\documentclass[pre,twocolumn,aps,showpacs,nofootinbib,superscriptaddress]{revtex4-1} % showkeys

\usepackage{tikz}

\usepackage{amsmath}
\usepackage{graphicx}% Include figure files
\usepackage{dcolumn}% Align table columns on decimal point
\usepackage{bm}% bold math
\usepackage{color}
\usepackage[normalem]{ulem}

\usepackage{amsfonts,epsfig,amsmath,amsthm,amssymb,graphics,verbatim}

\usepackage{color}
\newcommand{\be}{\begin{equation}}
\newcommand{\ee}{\end{equation}}
\newcommand{\bea}{\begin{eqnarray}}
\newcommand{\eea}{\end{eqnarray}}
\newcommand{\beann}{\begin{eqnarray*}}
\newcommand{\eeann}{\end{eqnarray*}}
\newcommand{\benn}{\begin{equation*}}
\newcommand{\eenn}{\end{equation*}}

\begin{document}

\title{Network topology near criticality in adaptive epidemics}

\author{Leonhard Horstmeyer}
\affiliation{Section for the Science of Complex Systems, CeMSIIS, Medical University of Vienna, 
Spitalgasse 23, A-1090, Vienna, Austria}
\affiliation{Complexity Science Hub Vienna, Josefst{\"a}dterstrasse 39, A-1090 Vienna, Austria}
\author{Christian Kuehn}
\affiliation{Faculty of Mathematics, Technical University of Munich, Boltzmannstr. 3, 85748 Garching M\"unchen, Germany}
\affiliation{Complexity Science Hub Vienna, Josefst{\"a}dterstrasse 39, A-1090 Vienna, Austria}
\author{Stefan Thurner}
\affiliation{Section for the Science of Complex Systems, CeMSIIS, Medical University of Vienna, 
Spitalgasse 23, A-1090, Vienna, Austria}
\affiliation{Complexity Science Hub Vienna, Josefst{\"a}dterstrasse 39, A-1090 Vienna, Austria}
\affiliation{Santa Fe Institute, 1399 Hyde Park Road, Santa Fe, NM 87501, USA}
\affiliation{IIASA, Schlossplatz 1, 2361 Laxenburg, Austria}

% \affiliation{%
%  Section for the Science of Complex Systems, CeMSIIS, Medical University of Vienna,\\ 
% Spitalgasse 23, A-1090, Vienna, Austria
% }%
% \author{Christian Kuehn}%
%  \email{Second.Author@institution.edu}
% \affiliation{%
% Faculty of Mathematics, Technical University of Munich, Boltzmannstr. 3, 85748 Garching M\"unchen, Germany\\
% }%
% \author{Stefan Thurner}
%  \altaffiliation[Also at]{Physics Department, XYZ University.}%Lines break automatically or can be forced with \\
% \affiliation{
%  Second institution and/or address\\
%  This line break forced% with \\
% }%
% \affiliation{
%  Third institution, the second for Charlie Author
% }%
% \author{Delta Author}
% \affiliation{%
%  Authors' institution and/or address\\
%  This line break forced with \textbackslash\textbackslash
% }%

% \author{Leonhard Horstmeyer$^{1}$, Christian Kuehn$^{2,3}$, and Stefan Thurner$^{1,3,4,5}$
% }
% %\email{stefan.thurner@meduniwien.ac.at} 
% 
% \affiliation{
% $^1$ Section for the Science of Complex Systems, CeMSIIS, Medical University of Vienna, 
% Spitalgasse 23, A-1090, Vienna, Austria\\
% $^2$ Faculty of Mathematics, Technical University of Munich, Boltzmannstr. 3, 85748 Garching M\"unchen, Germany\\
% $^3$ Complexity Science Hub Vienna, Josefst{\"a}dterstrasse 39, A-1090 Vienna, Austria \\
% $^4$ Santa Fe Institute, 1399 Hyde Park Road, Santa Fe, NM 87501, USA\\
% $^5$ IIASA, Schlossplatz 1, 2361 Laxenburg, Austria \\
% } 

\date{Version \today}

\begin{abstract}
We study structural changes of adaptive networks in the co-evolutionary 
susceptible-infected-susceptible (SIS) network model along its phase transition. 
We clarify to what extent these changes can be used as early-warning signs for the transition 
at the critical infection rate  $\lambda_c$ at which the network collapses and the system disintegrates. 
We analyze  the interplay between topology and node-state dynamics near criticality. 
Several network measures exhibit clear maxima or minima 
close to the critical threshold that could potentially serve as early-warning signs. 
These measures include the $SI$ link density, triplet densities, clustering, assortativity and the eigenvalue gap. 
For the $SI$ link density and triplet densities the maximum is found to originate from the co-existence of two power laws.
Other network quantities, such as the degree, the branching ratio, or the harmonic mean distance,
show scaling with a singularity at $\lambda=0$ and not at $\lambda_c$, 
which means that they are incapable of detecting the transition. 
\end{abstract}


\maketitle

%%%%%%%%%%%%%%%%%%%%%%%%%%%%%%%%%%%%%%%%%%%%%%%%%%%%%%%%%%%%%%%%%%%%%%%%%%%%%%%%%%%
\section{Introduction}
\label{sec:intro}
In recent years there has been an increasing focus on adaptive (or co-evolving) networks~\cite{GrossBlasius,GrossSayama,SzaboFath}. 
The essence of adaptive networks is that node-state dynamics influences the network topology -- and topology influences the node dynamics. 
Several adaptive network models have been phrased in the context of epidemics~\cite{GrossDLimaBlasius,ShawSchwartz}, 
game theory~\cite{PachecoTraulsenNowak,christoly2,ZimmermannEguiluzSanMiguel}, 
socio-dynamics~\cite{christoly1},
self-organized criticality~\cite{BornholdtRohlf,LevinaHerrmannGeisel}, 
financial markets~\cite{PolednaThurner}, 
and evolution~\cite{JainKrishna2}, just to name a few. 
These models are understood, either by simulations or by appropriate approximations, 
such as mean-field approximations and moment closure~\cite{DemirelVazquezBoehmeGross,christoly1,Gleeson1,KuehnMC}.

Adaptive networks show bifurcations or phase transitions, 
which means that they exist in at least two phases, one that is characterized by well-connected networks and another 
that has a drastically reduced link density (``collapsed'' phase).
The corresponding critical parameters separate the phases and can be computed explicitly for several models. 
It is known that for bifurcation-induced critical transitions, in the vicinity of these critical parameters (tipping points), 
so-called early-warning signs (precursor signals) exist that are linked to the phenomenon of critical slowing down, see e.g.~\cite{Wiesenfeld1}. 
For stochastic systems, slowing down can often be quantified by the autocorrelation and variance of the process.
In the context of adaptive networks critical slowing down is observed in terms of node properties \cite{KuehnZschalerGross,KuehnCT2}. 

A classic model for adaptive networks is the {\em co-evolving} SIS model, where the term co-evolving 
means that links and states--$S$ (susceptible) and $I$ (infected)--do not evolve independently. 
In the {\em static} SIS network model, where the network does not change over time, 
nodes are in a the $S$ or $I$ state. Each infected node recovers from infection at a rate $r$. 
An infected node can transmit the disease to connected susceptible nodes at a rate $\lambda$.  
In \cite{GrossDLimaBlasius} rewiring was introduced, where susceptible nodes may rewire a link from an infected node 
to a susceptible node at a rate $w$. This adaptive SIS model shows a different phase diagram, including a 
disease-free phase (almost all nodes $S$), an epidemic phase (almost all nodes $I$), a bi-stable phase,  
and an oscillatory phase~\cite{GrossDLimaBlasius,ShawSchwartz,GrossKevrekidis}. 
In the following, we focus on the phase transition \emph{from} the epidemic/bi-stable phase {\em to} the disease-free state. Depending on the context (infection, opinions, information etc.) the disease-free state can be have a postive, a negative or a neutral connotation. 
The transition happens at a critical infection rate, the so-called the \emph{persistence threshold}, $\lambda_c$.

The adaptive SIS model can be described with ``macroscopic equations'',  
where the stochastic node and link update dynamics is reduced to a system of ordinary differential equations 
(ODEs) that governs the fraction of infected nodes and the densities of the various link-types in the population. 
The equations are derived in the so-called heterogeneous pair approximation (PA).  
It is possible to estimate the critical infection rate at the persistence threshold. 
We denote the fraction of infected nodes by $\rho=[I]/N$, where $N$ is the number of nodes. 
The per-node density of $SS$ links, $SI$ links and 
$II$ links are denoted by 
$\rho_{SS}=[SS]/N$, $\rho_{SI}=[SI]/N$, and $\rho_{II}=[II]/N$, respectively. 
We also consider the densities of the motives,   
$\rho_{SSI}= [SSI]/N$ and $\rho_{ISI}=[ISI]/N$, 
which denote the respective triplet density per node. 
These densities are random variables, however, we denote their expectation values with the same variables. 
The evolution equations for the expectation values (up to second order) are given by~\cite{GrossDLimaBlasius},
\begin{subequations}
\label{eq:PAs}
\begin{align}
\frac{d\rho}{dt}&=\lambda \rho_{SI}-r\rho\label{eq:PA1}\\
\frac{d\rho_{II}}{dt}&=\lambda \rho_{SI}+\lambda \rho_{ISI}-2r\rho_{II}\label{eq:PA2}\\
\frac{d\rho_{SS}}{dt}&=(r+w)\rho_{SI}-\lambda \rho_{SSI}\label{eq:PA3} \quad . 
\end{align}
\end{subequations}
Let $\langle k \rangle$ denote the average degree and note, that since the total link density, 
\be
\label{eq:masscon}
\rho_{SS}+\rho_{SI}+\rho_{II}=\frac{\langle k \rangle}{2}
\ee
is conserved in the rewiring process, the seemingly missing $\rho_{SI}$-equation can be eliminated.  
Equations~\eqref{eq:PA2}-\eqref{eq:PA3} are not closed because they depend on triplet densities. 
To close them, one can use e.g. the homogeneous pair approximation\footnote{The quality of this 
approximation can be checked in simulations, see Appendix \ref{ssc:PA}.} 
that neglects correlations between links, 
\be
\label{eq:tripletapprox}
\rho_{SSI}\approx 2\frac{\rho_{SI}\rho_{SS}}{1-\rho},\qquad 
\rho_{ISI}\approx \frac{\rho_{SI}\rho_{SI}}{1-\rho} \quad .
\ee
One can now solve for the stationary solution of the PA. 
The disease-free state is always a steady state, but looses 
stability at the so-called invasion threshold, for which the PA yields,  
$\lambda^{\text{invasion}}=(r+w)/\langle k\rangle$. 
In the PA the endemic state starts to be stable at the persistence threshold,
\begin{equation}
\lambda_c =\frac{2r}{\mu^2}\left(\sqrt{1+\frac{w\mu^2}{r}}-1\right) \quad ,
\label{eq:persistence_threshold}
\end{equation}
where we define $\mu:=\langle k \rangle -1$ as the approximate average excess degree. 
For $r\ll w\mu^2$, we have $\lambda_c \approx 2\sqrt{wr}/\mu$.
\begin{figure}
\centering
% \hspace{-10ex}
\begin{tikzpicture}
% \node (empty) at (0,0) {};
\node[inner sep=0pt] (Fig1) at (0,0)
{\includegraphics[height=4.15cm]{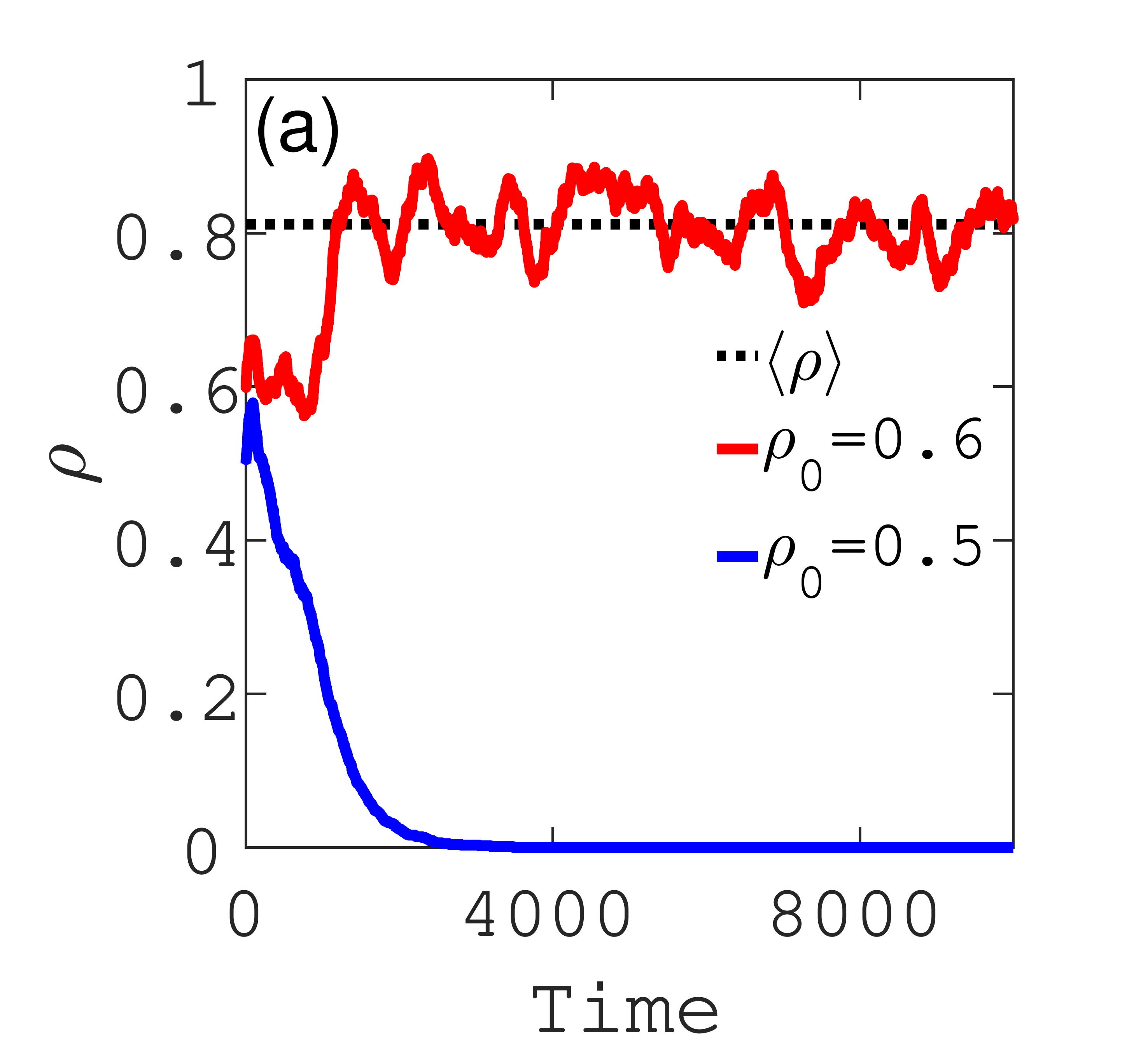}};
\node[inner sep=0pt] (Fig2) at (4.15,0)
{\includegraphics[height=4.15cm]{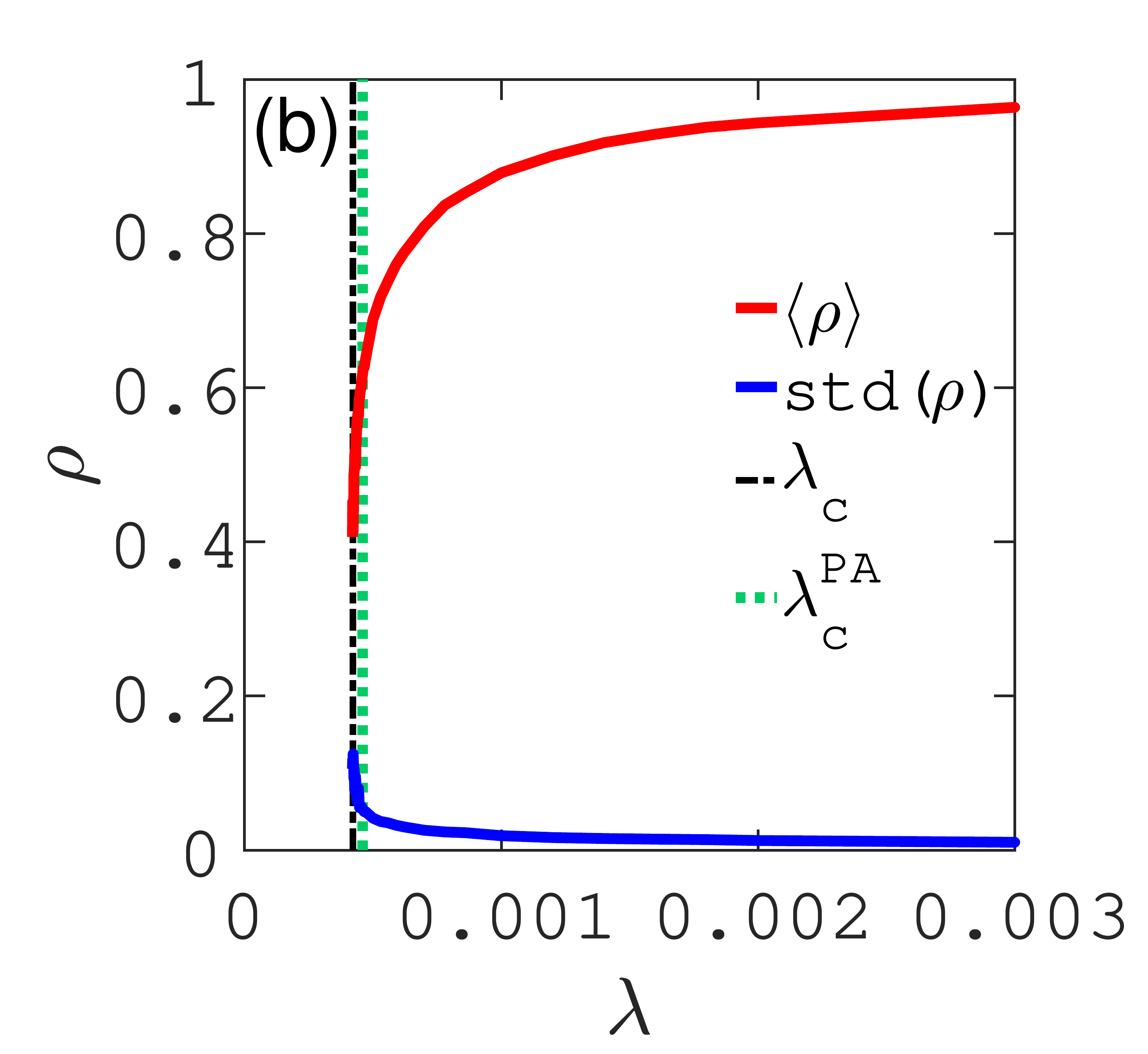}};
\end{tikzpicture}
 \caption{(a) Disease prevalence in the adaptive SIS model. Two timeseries are shown with 
	an initial prevalence of 60\% (red) and 50\% (blue). The network 
	is initialized as an Erd\"os R\'enyi graph of size $N=400$ with an average 
	degree of $\langle k \rangle=20$ and the remaining parameters are $r=0.002$, $\lambda=0.001$, and $w=0.01$. The dashed line indicates the stationary value in the epidemic state. 
	(b) Stationary endemic prevalence as a function of infection rates (red) 
	and its standard deviation (blue) are shown for the same parameters. The numerical (green, dotted) and pair approximation (black, dashed) values of the persistence threshold are also indicated. Their values for these parameters are $\lambda_c=4.2\times 10^{-4} (\pm 0.2\times 10^{-4})$ and $\lambda_c^{PA}=4.6\times 10^{-4}$ (c.f. Equation \eqref{eq:persistence_threshold}) respectively . 
\label{fg:Prevalence}
}
\end{figure}
Figure~\ref{fg:Prevalence}(a) shows two timeseries of the prevalence $\rho$ in a simulation of the adaptive SIS model in the bistable 
regime. For the specific initial conditions shown, the dynamics either enters the stationary endemic state or 
the disease-free state. The smaller the initial disease prevalence, the higher the probability of ending 
up in the disease-free state. If the system is in the endemic state, it explores its phase space stochastically. 
In Fig. \ref{fg:Prevalence}(b) we show the average prevalence as a function of the infection rate $\lambda$. 
It asymptotically approaches $1$ for large $\lambda$. 
Close to $\lambda_c$, the prevalence decreases sharply and eventually the endemic state ceases to exist. 
The standard deviation of $\rho$ increases as the infection rate approaches $\lambda_c$ from above~\cite{KuehnCT2}. 
This reflects that around the critical point, fluctuations become larger--the chance of an extinction event increases. 
Setting the right-hand-side of \eqref{eq:PAs} to zero and solving for $\rho$ yields the equilibrium curve for the prevalence in the PA 
and shows the leading order $\rho\propto (\lambda-\lambda_c)^{\frac{1}{2}}$ behaviour, 
\begin{equation}
 \rho=
 1-\frac{\lambda\mu}{2(w-\lambda)}+ \frac{\sqrt{\lambda^2\mu^2-4r(w-\lambda)}}
 {2(w-\lambda)}
\end{equation}
Note, that the singularity at $\lambda=w$ is removable by assigning $\rho(\lambda=w)=1-\mu / w -2r / (w\mu)+\mu$.

\begin{figure*}
\centering
% \hspace{-10ex}
\begin{tikzpicture}
\node[inner sep=0pt] (Fig1) at (0,0)
{\includegraphics[height=4.15cm]{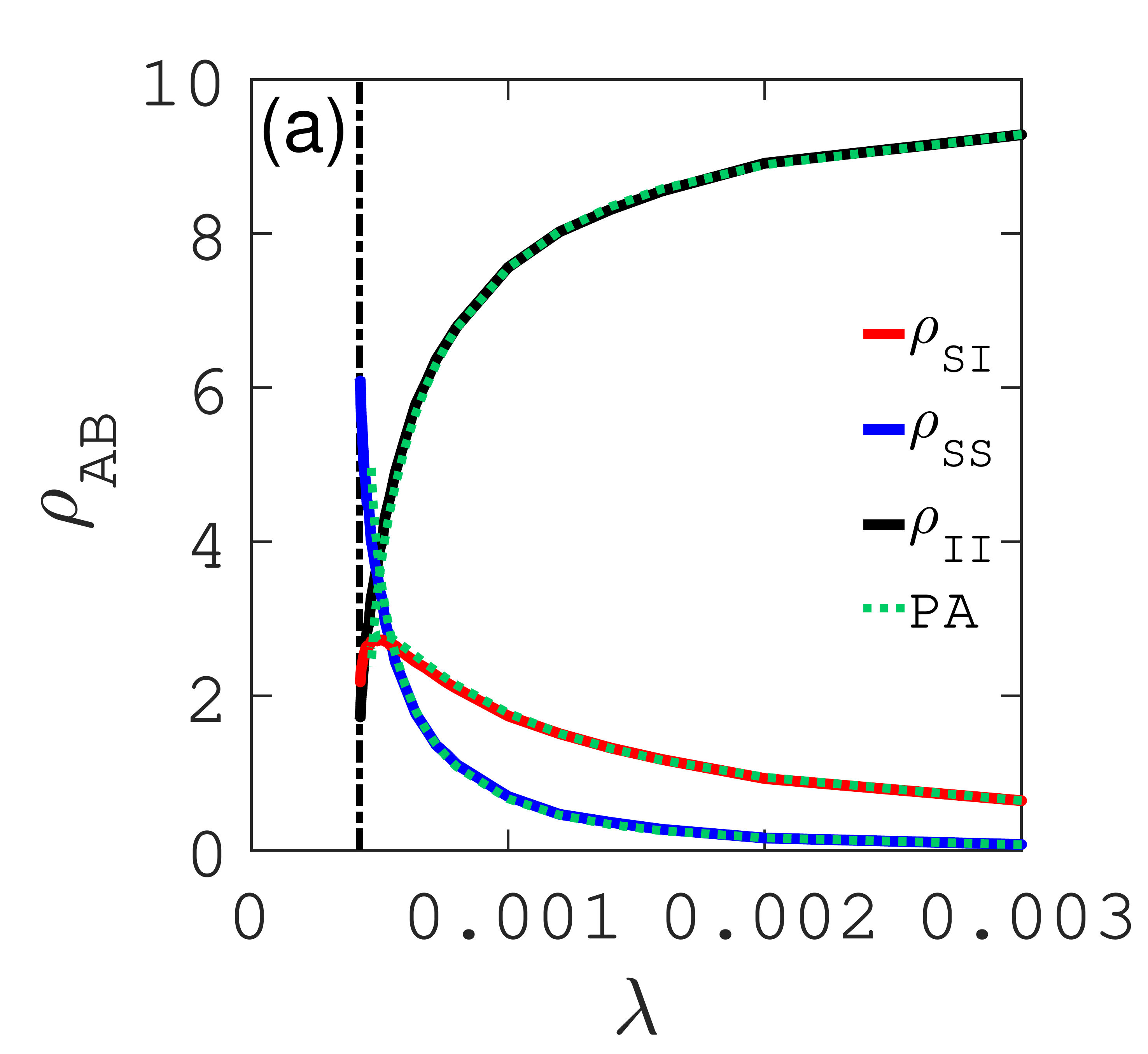}};
\node[inner sep=0pt] (Fig2) at (4.5,0)
{\includegraphics[height=4.15cm]{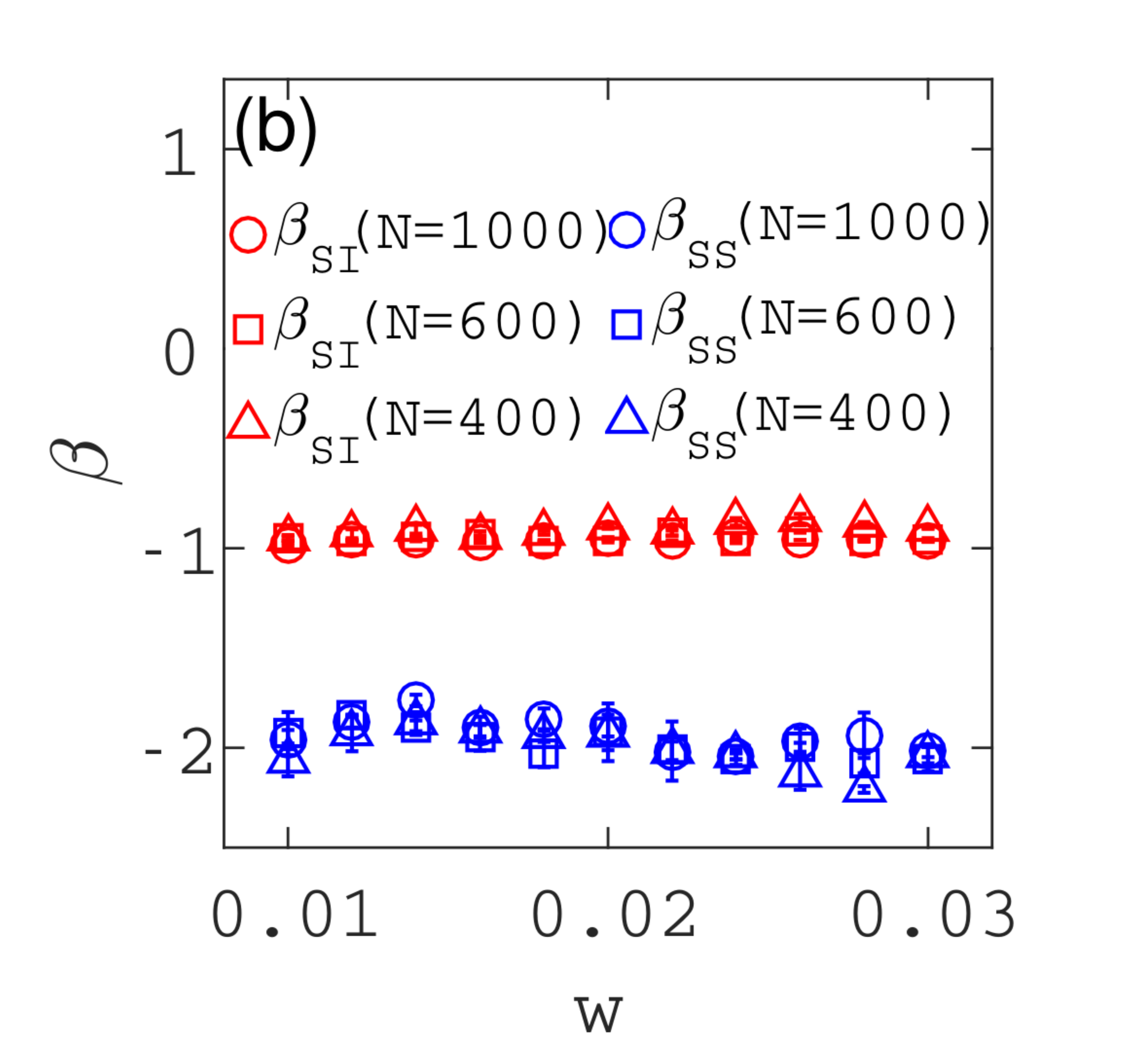}};
\node[inner sep=0pt] (Fig3) at (8.9,0)
{\includegraphics[height=4.15cm]{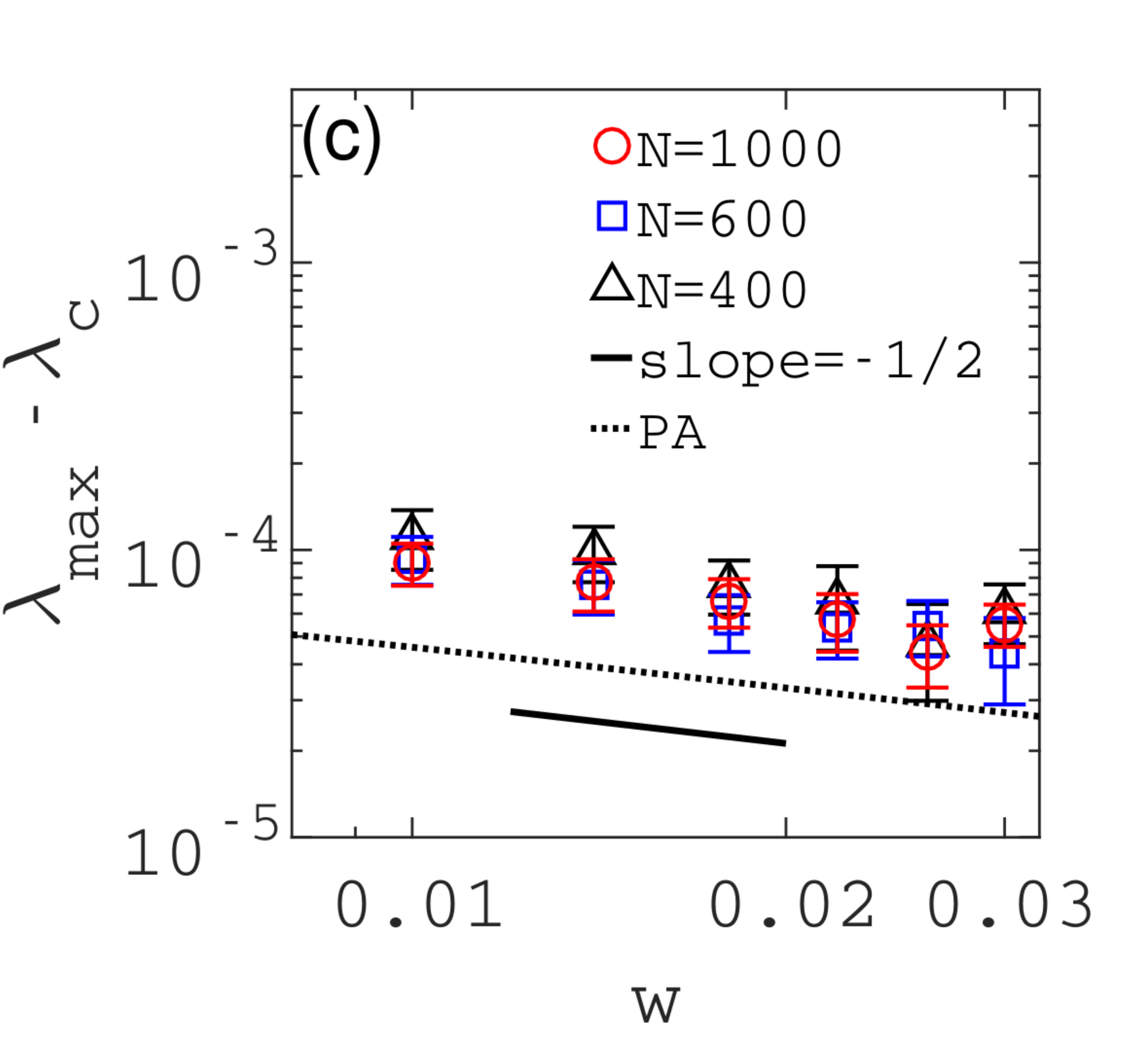}};
\node[inner sep=0pt] (Fig4) at (13.5,0)
{\includegraphics[height=4.15cm]{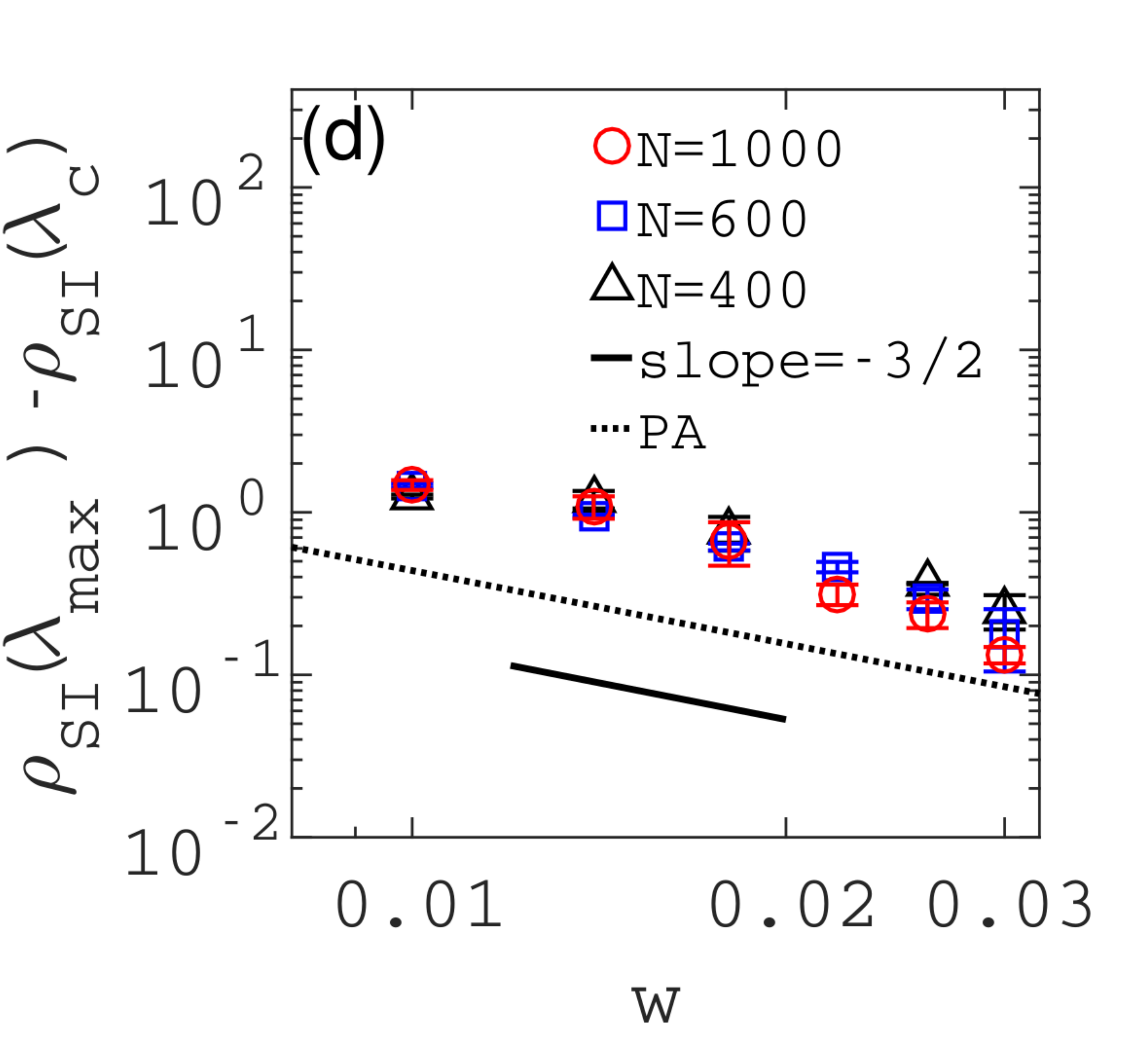}};
\end{tikzpicture}
  \caption{(a) Critical curves of link densities are shown for numerical simulations (dashed) and in the PA (dotted). 
$SI$ links (red), $SS$ links (blue) and $II$ links (black). $r=0.002$, $w=0.01$, $\langle k \rangle=20$, and $N=400$. 
	(b) Exponents of the tails are fitted to $\sim \lambda^{\beta}$ for $SI$ links (red), and $SS$ links (blue) as a function of 
	rewiring rates $w$ and system sizes. 
	(c) Location and 
	(d) size of the maxima in the $SI$ link density as a function of $w$ and system sizes, and for reference, also their values in the PA.
\label{fg:linkdensities}
}
\end{figure*}

At this level the adaptive $SIS$ network model is well-understood.
In this paper we ask, how critical transitions in adaptive network dynamics
are reflected in the network topology of the underlying network(s). 
The practical motivation behind this question is whether it is possible to use the monitoring of 
the networks to infer the closeness to the critical point of adaptive systems. 
We are interested to what extent early-warning signals can be derived from 
eventual rearrangements of network structures close to the critical transition $\lambda_c$.
In particular we ask, whether the networks' structural changes follow certain scaling laws and if 
those can be used for predicting the upcoming transition. 
We take a first step in this direction by studying the adaptive network model proposed in \cite{GrossDLimaBlasius}.
Our main results are:

\begin{itemize}
\item[(R1)] For several network-related quantities, $SI$ link densities, triplet densities, clustering, assortativity, and the eigenvalue gap, 
there exists a specific crossover of two scaling laws near criticality. 
As a consequence, these quantities show local extrema close to the persistence threshold. 
This effect can be explained within the PA framework. 
These extrema might indeed serve as potential candidates for network-based early-warning signs. 
The eigenvalue gap might be an especially practical measure. 
\item[(R2)] Some network-related quantities, such as the degree, the effective branching ratio, and the harmonic mean distance,  
behave as if there existed a critical point that has a singularity located at the origin $\lambda^{\rm network}_c \sim 0$. 
Their critical curves end abruptly at the threshold $\lambda_c$, so that their role as early-warning signs is limited. 
 \item[(R3)] Fluctuations and correlations increase for topological measures near $\lambda_c$, which might 
be an additional signal, when approaching the tipping point. 
\end{itemize}

In summary, we show that topological changes in adaptive networks close to the critical point carry  
potential information to improve predictability of critical transitions through early-warning signs. 
In this sense the information that is neglected in the coarse-graining approach does indeed contain a crucial layer of structural information.

%%%%%%%%%%%%%%%%%%%%%%%%%%%%%%%%%%%%%%%%%%%%%%%%%%%%%%%%%%%%%%%%%%%%%%%%%%%%%%%%%%%%%%%%%%%%%%%
\section{The critical network}
\label{sc:critical}

We present the results of a numerical study of network properties near the persistence threshold. 
We employ the Gillespie algorithm for the simulation, which samples the stochastic process in an unbiased way \cite{Gillespie1}. 
For infection rates close to the threshold, we use the quasi-stationary method, which was used in \cite{DeOliveiraDickman},  
and applied to epidemic networks \cite{FerreiraFerreiraPastor,FerreiraCastellanoPastor}.
In this paper we focus on link densities of various link types, triplet densities, the effective 
branching ratio, the clustering coefficient, degree distribution, degree assortativity, compactness, and finally, 
spectral properties of the adjacency matrix.

%%%%%%%%%%%%%%%%%%%%%%%%%%%%%%%%%%%%%%%%%%%%%%%%%%%%%%%%%%%%%%%%%%%%%%%%%%%%%%%%%%%%%%%%%%
\subsection{Link densities}
\label{ssc:linkdensities}

The densities of $SS$, $SI$ and $II$ links reveal a detailed picture of the 
mechanisms that are at work near the critical persistence threshold. 
In Fig.~\ref{fg:linkdensities} (a) we show the average per-node densities for $SS$, $SI$ and $II$ links in the endemic
stationary state for a range of infection rates near the persistence threshold. 
$SS$ and $SI$ link densities approach $0$ asymptotically for large infection rates because rewiring cannot keep up with the infections.
Hence $II$ links dominate that regime. 

Close to the persistence threshold the density of $SS$ links ($II$ links) increase (decrease). 
For the $SI$ links, however, there is a distinctive maximum that deserves attention. 
One can express this observation in terms of the derivatives 
with respect to the infection rate. Using Eq.~\eqref{eq:masscon}, we have,  
$\rho_{SI}^{\prime}=-\rho_{SS}^{\prime}-\rho_{II}^{\prime}$, 
where $\rho_{AB}^{\prime}$ denotes the rate of change of the $AB$ link density.
Thus for infection rates near the threshold the $SS$ link density must decrease faster 
than $II$ links increase; for slightly higher infection rates the roles interchange. 
So we conclude that $SS$ and $II$ links scale differently near the threshold, as can be seen in 
Fig.~\ref{fg:linkdensities} (b). The tail of $\rho_{SS}$ scales roughly as $\lambda^{-2}$, and the tails of $\rho_{SI}$ and, 
by link conservation,  $\rho_{II}$ scale as $\lambda^{-1}$. 
The exponent for $\rho_{SI}$ is systematically slightly overestimated, 
because the square-root behavior interfears slightly. This behavior is robust with respect to system size.

Using the PA in Eq.~\eqref{eq:PAs} we get the following estimate,  
\begin{equation}
\rho_{SI}= 
\frac{r\mu}{2(w-\lambda)}
\left[\sqrt{1-\frac{4r(w-\lambda)}{\lambda^2\mu^2}}-
1\right]+\frac{r}{\lambda} \quad . 
\label{eq:pa_silinks}
\end{equation}
Note, that the singularity at $\lambda=w$ is again removable. 
The functional form close to the critical point is given by 
\begin{equation}
\rho_{SI}
= \frac{r}{\lambda_c} + \frac{ (\lambda_c-2)  r\mu \sqrt{\frac{\lambda_c}{2} + \frac{r}{\mu^2}}}{2 \lambda_c (w-\lambda_c)} 
\sqrt{\Delta \lambda}%\nonumber \\& 
+\mathcal O(\Delta \lambda) \quad , 
\label{eq:pa_silinks_crit}
\end{equation}
where $\Delta \lambda=\lambda-\lambda_c$. Obviously, the density of $SI$ links 
follows a square-root behavior near the critical point with a positive slope, 
as expected from the universal behaviour of the fold bifurcation~\cite{GH} that is present at this point~\cite{GrossDLimaBlasius}. 
For larger infection rates, $\lambda\gg\lambda_c$ the PA predicts a decay that is dominated by $\lambda^{-1}$. 
Therefore a maximum must occur in between. 
There are two (critical) exponents of the $SI$ link densities. 
In the vicinity to the threshold we expect a square-root behavior $\rho_{SI} \sim \Delta \lambda^{\frac12}$, 
for larger $\lambda$ we get a power law decay with an exponent $\rho_{SI} \sim (\lambda-0)^{-1}$. 
Effectively, we can write the result obtained in Eq.~\eqref{eq:pa_silinks} in the functional form, 
\begin{equation}
f(\lambda) =  \alpha \lambda^{-\frac{3}{2}} (\Delta\lambda)^{\frac{1}{2}}+ \beta \lambda^{-1} + f_0 \quad .  
\label{eq:model}
\end{equation}
The equation has three regimes. For $\Delta \lambda$ much smaller than $\lambda_c$, 
$f(\lambda) \approx \gamma(\Delta \lambda)^{1/2}+\delta$, with $\gamma=\alpha\lambda_c^{-3/2}$ 
and $\delta=\beta / \lambda_c+f_0$. 
When we identify $\rho_{SI}$ with $f$ we can solve for $\alpha$, $\beta$ and $f_0$, using Eq. \eqref{eq:pa_silinks_crit}. 
For $\lambda \gg \lambda_c$ it behaves as  $f (\lambda) \approx \zeta \lambda^{-1}-\xi \lambda^{-2}+f_0$ 
with $\zeta=\alpha+\beta$ and $\xi=\alpha \lambda_c / 2$, which can be checked by a Taylor expansion of Eq. \eqref{eq:model}. 
When we identify $\rho_{SI}$ with $f$, we obtain $\zeta=r$ and $\xi=r^2 / \mu$, again with an expansion of Eq. \eqref{eq:pa_silinks} 
for large $\lambda$. One can then solve for $\alpha$ and $\beta$. 
The intermediate regime contains the maximum. 
In Fig. \ref{fg:linkdensities} (c) and (d) we investigate the location and hight of the maximum of $\rho_{SI}$ with respect to the rewiring rate $w$. 
Both, the distance of the maximum from the threshold and its size seem to follow a power laws $w^{-1/2}$ and $w^{-3/2}$, respectively. 
This behavior is also seen in the PA which is shown for comparison. 
Since the PA becomes less accurate towards the threshold, it is not surprising that the PA estimates differ in absolute terms but not qualitatively (slope). The behavior is again robust with respect to system size.

%%%%%%%%%%%%%%%%%%%%%%%%%%%%%%%%%%%%%%%%%%%%%%%%%%%%%%%%%%%%%%%%%%%%%%%%%%%%%%%%%%%%%%%%%%
\subsection{Triplet densities}
\label{ssc:triplets}

\begin{figure}
\centering
\begin{tikzpicture}
\node[inner sep=0pt] (Fig1) at (0,0)
{\includegraphics[height=4.15cm]{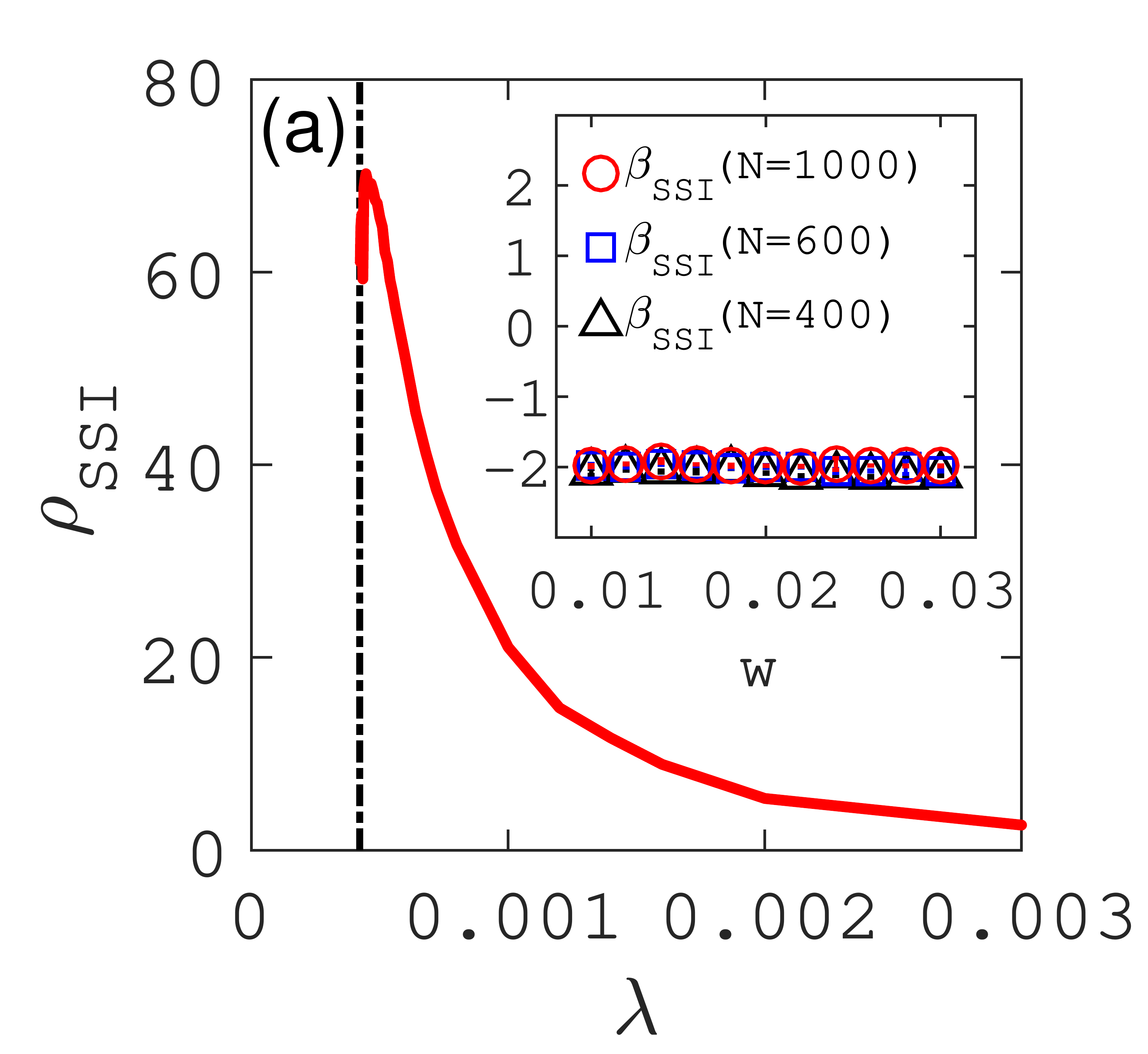}};
\node[inner sep=0pt] (Fig2) at (4.3,0)
{\includegraphics[height=4.15cm]{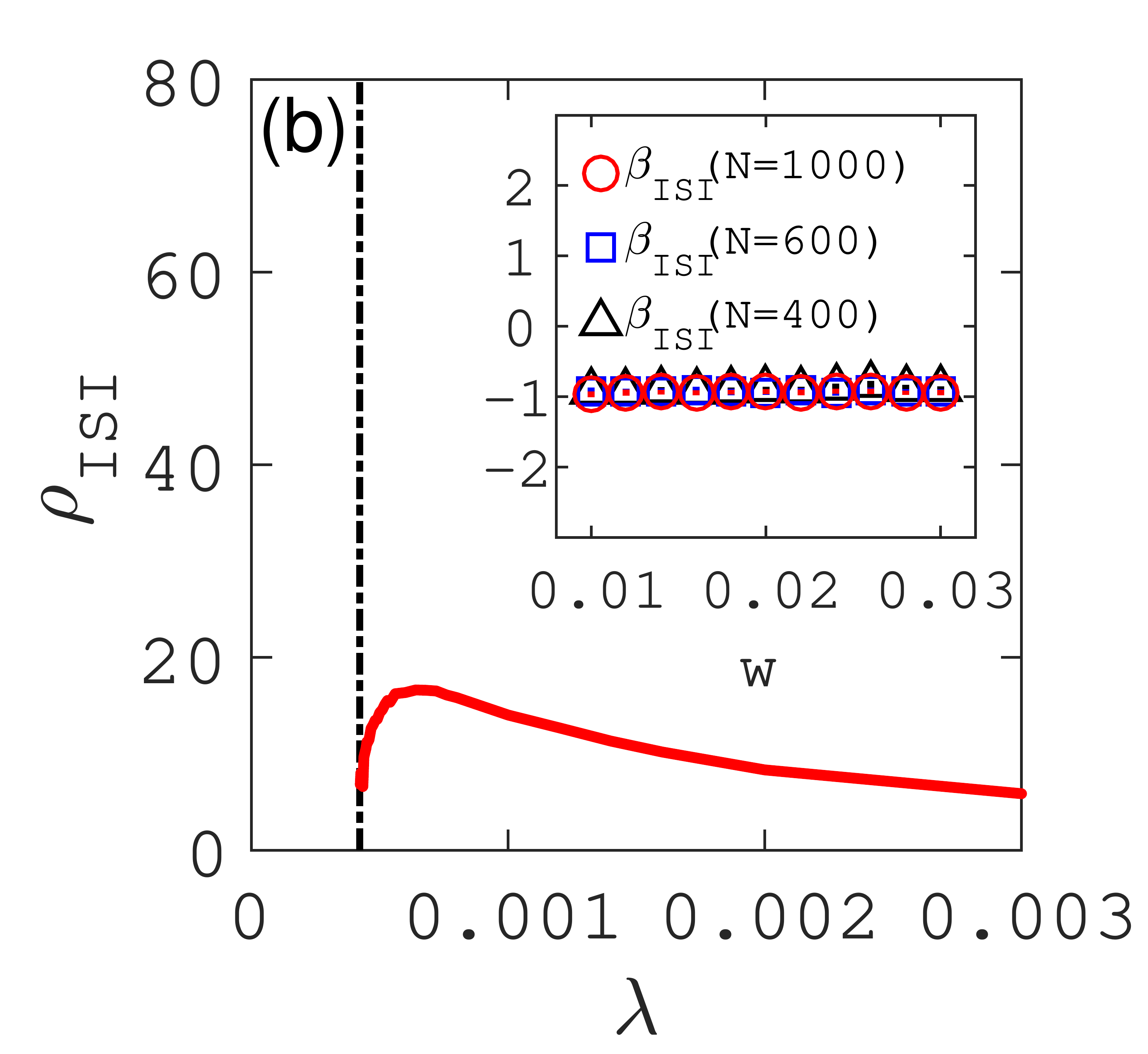}};
\end{tikzpicture}
 \caption{(a) Density of $SSI$ triplets.
	The inset shows the critical exponents of the tail for a range of rewiring rates and system sizes. It is about $-2$.
	(b) $ISI$ triplet density. The inset shows a critical exponent of $-1$ irrespective of $w$ and $N$. Parameters as before.
	\label{fg:triplets}
}
\end{figure}

We now focus on triplets where the central node is susceptible. 
In particular, we study the $SSI$ and $ISI$ motives that are crucial in the PA in Eq.~\eqref{eq:tripletapprox}.
Figure \ref{fg:triplets} shows the per-node densities for the $SSI$ and $ISI$ triplets. 
As for the $SI$ links, one can distinguish three regimes: the asymptotic regime of large infection rates, 
the critical regime of infection rates very close to $\lambda_c$ and a mid-range regime, containing a maximum. 
The maximum of the $ISI$ triplets is further away from the threshold than in the $SI$ link case. 
This can be understood in the PA, Eq.~\eqref{eq:tripletapprox}. 
The density of $ISI$ triplets is approximately the square of the $SI$ link densities, divided by the fraction 
of susceptible nodes. The square of a function will not change the position of its maximum, however, division will. 
Consider,  
\benn
\left(\frac{\rho^2_{SI}}{\rho_S}\right)^{\prime}=
\frac{\left(\rho_{SI}^2\right)^{\prime}}{\rho_S}-
\left(\frac{\rho_{SI}}{\rho_S}\right)^2(\rho_S)^{\prime}
\eenn 
This expression is positive for the infection rate, where $\rho_{SI}$ becomes maximal  
since the first term vanishes and the second is positive (because of the decrement of the susceptible density). 
Therefore the maximum of $\rho_{ISI}$ has not yet been attained at this rate. 
A similar analysis can be done for the $SSI$ triplet.

%%%%%%%%%%%%%%%%%%%%%%%%%%%%%%%%%%%%%%%%%%%%%%%%%%%%%%%%%%%%%%%%%%%%%%%%%%%%%%%%%%%%%%%%%%%%
\subsection{Effective branching ratio}
\label{ssc:ebr}

\begin{figure}
\centering
% \hspace{-10ex}
\begin{tikzpicture}
\node[inner sep=0pt] (Fig1) at (0,0)
{\includegraphics[height=4.15cm]{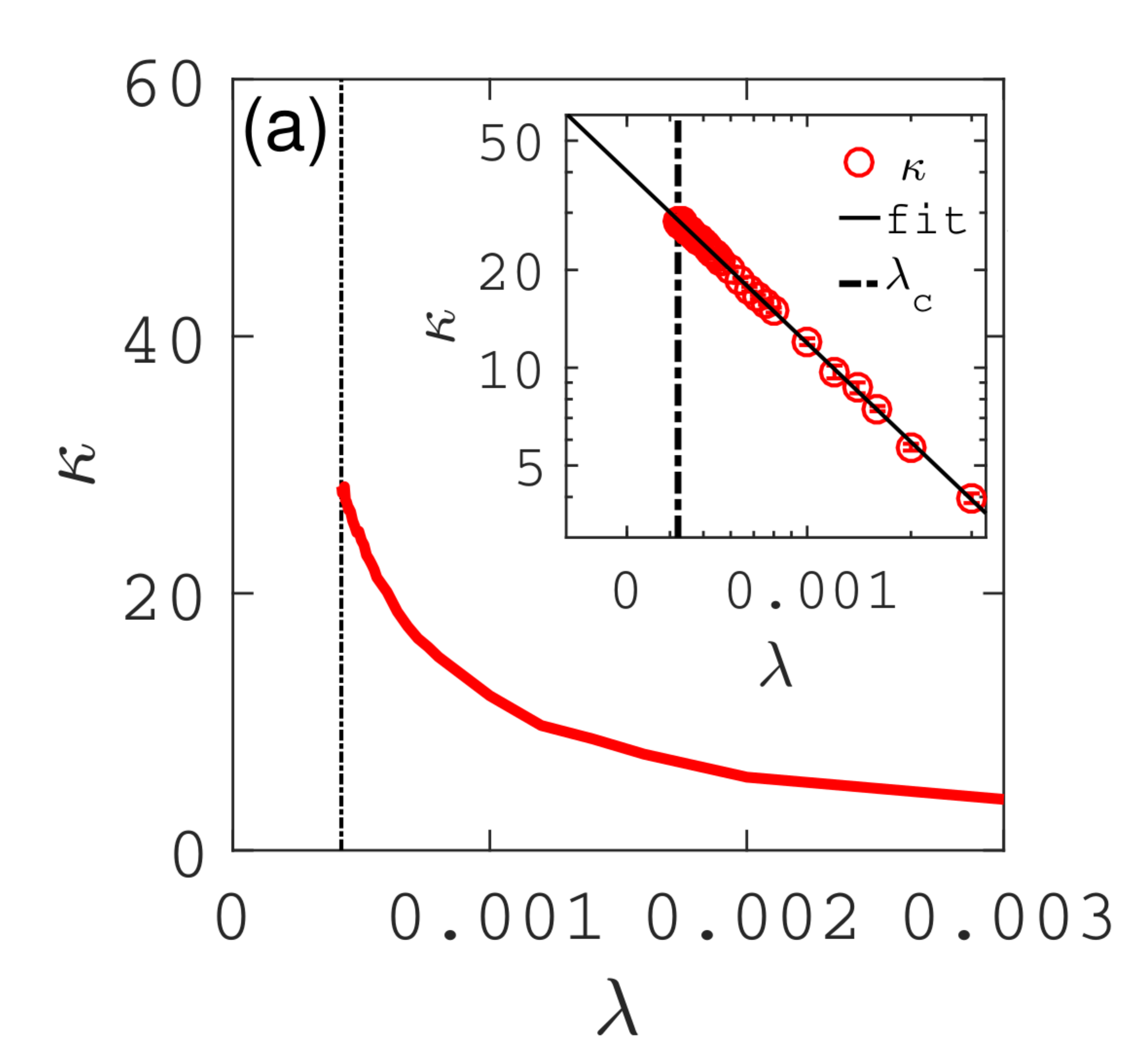}};
\node[inner sep=0pt] (Fig2) at (4.5,0)
{\includegraphics[height=4.15cm]{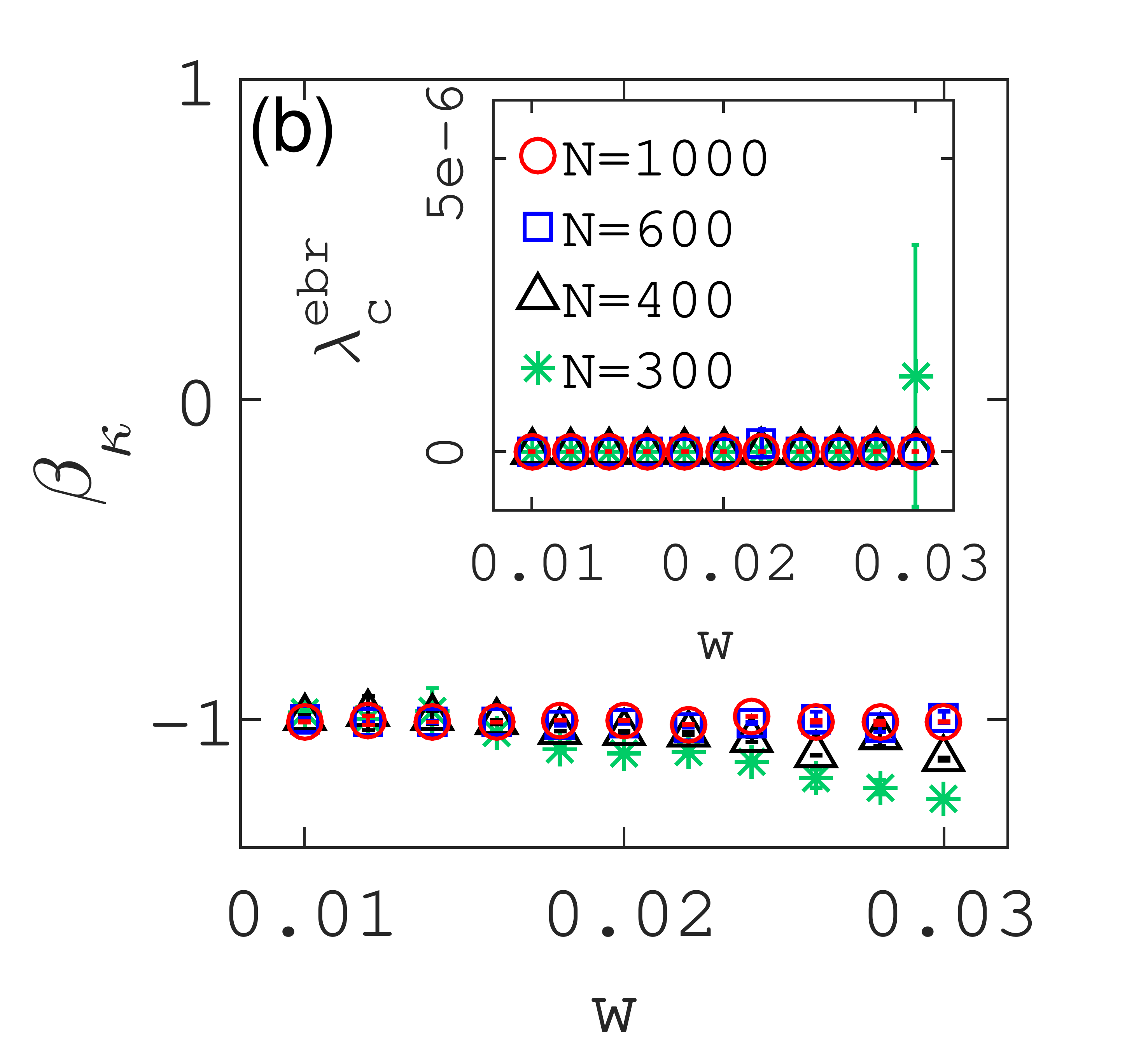}};
\end{tikzpicture}
    \caption{(a) Effective branching ratio $\kappa=SSI/SI$ (red). 
    		The inset is a log-log plot with a fitted slope, $\beta_\kappa = -1.01$. Parameters as before. 
		(b) Exponent $\beta_\kappa$ for a range of rewiring rates and system sizes. 
		The inset shows the fitted threshold $\lambda_c^{\text{ebr}}$. 
}
\label{fg:EBR}
\end{figure}

The effective branching ratio is defined as 
\begin{equation}
\kappa=\frac{[SSI]}{[SI]}=\frac{\rho_{SSI}}{\rho_{SI}} \quad , 
\end{equation}
and quantifies the number of potential secondary infections for a given primary infection. 
Figure~\ref{fg:EBR}(a) shows $\kappa$ in log-log scale. 
The effective branching ratio does not have a maximum but follows a power law with an exponent $\alpha \approx -1$. 
The power law is clearly of the form, $\kappa\propto (\lambda-\lambda^{\rm ebr}_c)^{-\alpha}$, 
where $\lambda^{\rm ebr}_c  \approx 0$. 
The critical transition at $\lambda_c$ is not detected by the effective branching ratio.

\begin{figure}[t]
\centering
% \hspace{-10ex}
\begin{tikzpicture}
% \node (empty) at (0,0) {};
\node[inner sep=0pt] (Fig1) at (0,0)
{\includegraphics[height=4.15cm]{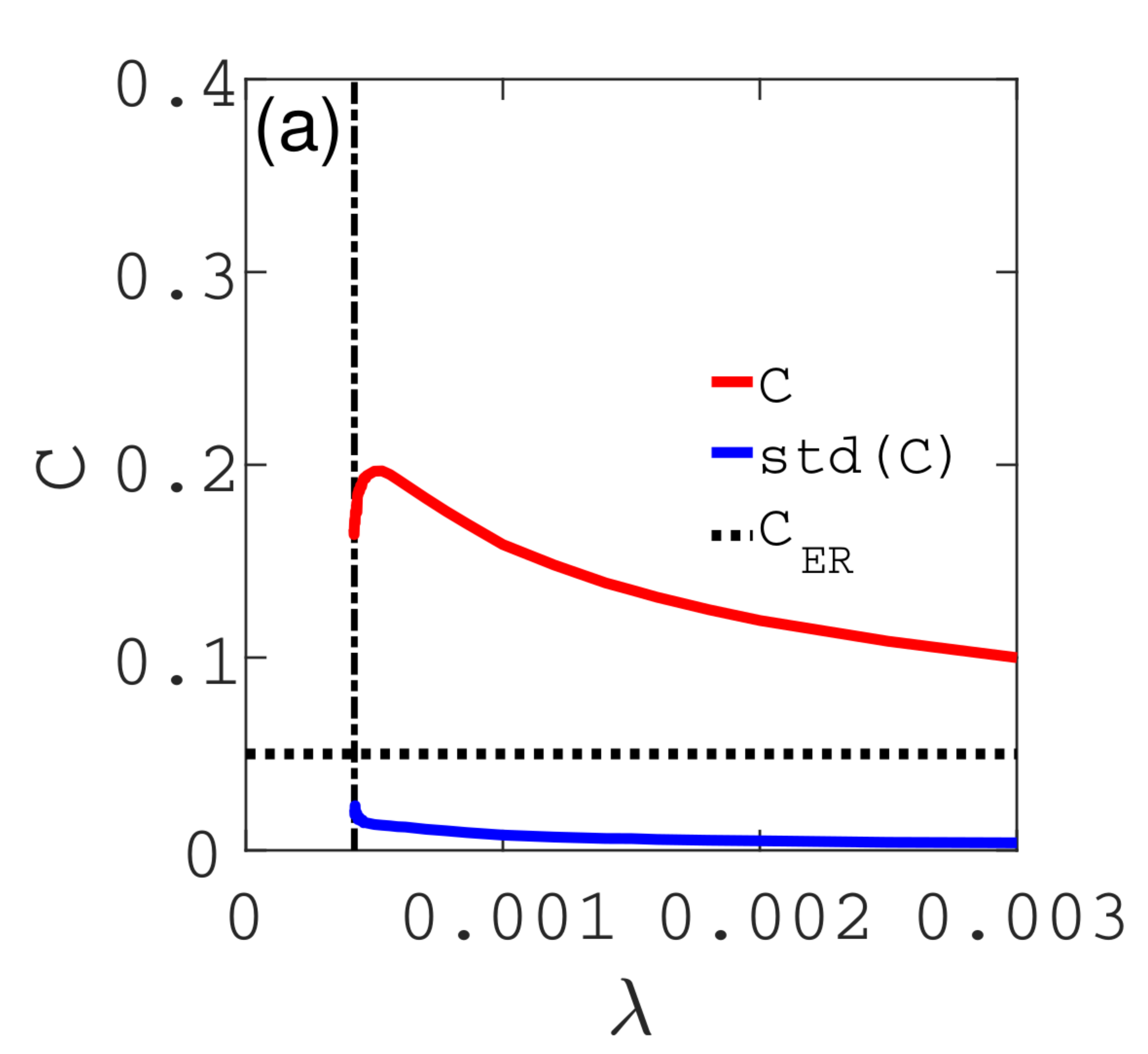}};
\node[inner sep=0pt] (Fig2) at (4.5,0)
% {\includegraphics[width=0.251\linewidth]{finalpics/average_local_clustering_gimped.pdf}};
{\includegraphics[height=4.15cm]{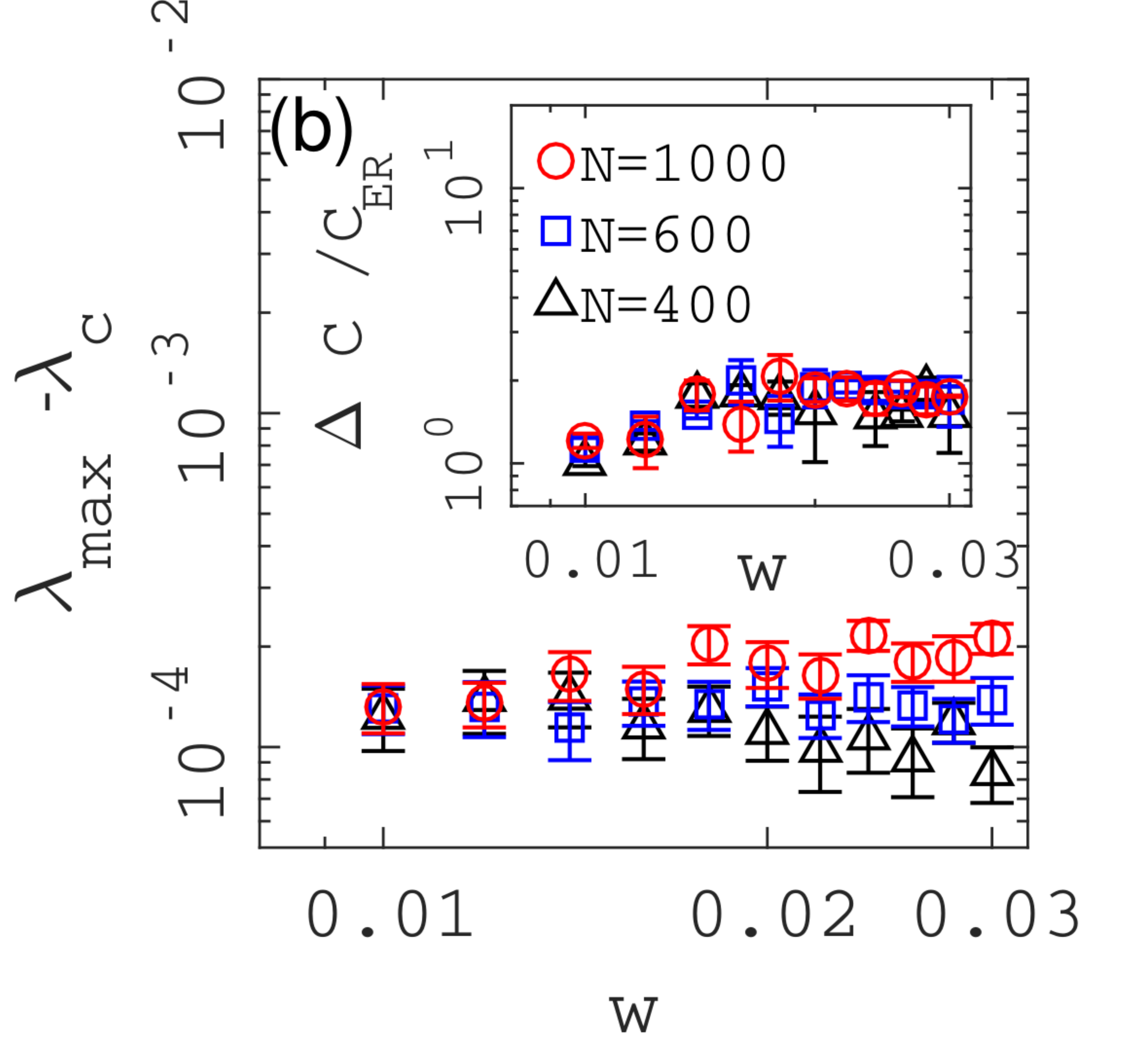}};
\end{tikzpicture}
 \caption{(a) Clustering coefficient (red), its standard deviation around the equilibrium state (blue), 
  	and the clustering coefficient of the Erd\"os R\'enyi graph (dotted) for $N=400$, $\langle k\rangle=20$, $r=0.002$, and $w=0.01$. 
	(b) Distance and relative size (inset) of the local maximum for various $w$ and $N$. 
	The infection rate at which the critical curve becomes maximal is $\lambda_{\text{max}}$. 
	Since the clustering coefficient vanishes as $N\to \infty$ at constant link density, 
	we rescale by $C_{ER}=\langle k \rangle / N$. Parameters are otherwise the same.
\label{fg:globalclust}
	}
\end{figure}

We measure the exponent $\beta_\kappa$ for various rewiring rates $w$ and system sizes $N$ in Fig.~\ref{fg:EBR}(b). 
For finite system sizes, we find that $\beta_\kappa$ decreases roughly linearly with the rewiring rate. 
With larger system size this dependence becomes weaker. 
Hence, we infer that the density of $SSI$ triplets (Fig.~\ref{fg:triplets}(a)) is just the product of the 
$SI$ link density and a power law with exponent $-1$. For secondary infections we 
conclude that the risk is highest just at the threshold, 
even though the risk of an initial infection--indicated by the $SI$ link density--is not maximal at the critical point. 

%%%%%%%%%%%%%%%%%%%%%%%%%%%%%%%%%%%%%%%%%%%%%%%%%%%%%%%%%%%%%%%%%%%%%%%%%%%%%%%%%%%%%%%%%%%%%%%

\subsection{Clustering coefficient}
\label{ssc:cc}

The clustering coefficient $C$ measures the number of closed triangles with respect to the total number of triangles in the entire network. 
It is given in terms of the adjacency matrix $A$ of the network by 
\begin{equation}
C=\frac{Tr A^3}{\sum_{ij} (A^2)_{ij} - Tr (A^2) }\;. \label{eq:clcoeff}
\end{equation}
Figure~\ref{fg:globalclust}(a) shows the clustering coefficient %~\eqref{eq:clcoeff} 
near $\lambda_c$. The qualitative behavior is again similar to the $SI$ link density, or the $SSI$ and $ISI$ triplet densities. 
Since there are almost no susceptible nodes in the regime of large infection rates, 
the stationary network behaves like an Erd\"os R\'enyi graph, whose clustering coefficient is given by 
$C_{\rm ER}=\langle k \rangle/N$, 
where $\langle k\rangle =1/N \sum_{i}k_i=2L / N$, $L$ is the total number of links. 
$C_{ER}$ is the limiting value for large infection rates (dotted horizontal line). 
Rewiring creates and destroys triangles. The clustering coefficient depends on the average net effect. 
The appearance of the maximum can be explained by this net effect in the three regimes. 
For high infection rates, a rewiring event has a much higher chance of closing an open triangle rather than destroying one, 
due to the high connectivity of the susceptible graph. 
However, the number of rewireable links is very low, which results in an asymptotically vanishing net effect. 
For infection rates very close to $\lambda_c$, there are many more susceptible nodes, and 
the chance for creating a closed triangle is only slightly larger than the chance to destroy one. 
The average net effect is nevertheless present. 
The largest effect occurs approximately there, where the number of $SI$ links, and hence the total rewiring rate, is maximal.

Fitting a power law to the tail of $C$ is sensitive to the interval choice and whether the ER limiting value is enforced or not. 
Parameter values of the fitted exponent vary depending on these choices. Since the clustering coefficient is a nonlinear function of graph motives, it is likely that multiple power laws of the respective motives interfere, which leads to the aforementioned sensitivity.

We denote by $\lambda_{\text{max}}$ the infection rate at which the critical curve becomes maximal. 
In Fig.~\ref{fg:globalclust}(b) we show the distance of $\lambda_{\text{max}}$ from the threshold $\lambda_c$ as a function of the rewiring rate. 
This distance does not decline for all system sizes $N$, as it is the case for the maxima of $\rho_{SI}$. 
The trend is rather that the distance rises for larger $N$.
The size of the maximum $\Delta C=C(\lambda_{\text{max}})-C(\lambda_c)$ does not decline as a function of $w$, but levels out. 
However, $\Delta C\propto N^{-1}$, which can be seen from the inset of Fig.~\ref{fg:globalclust}(b), 
where we rescale $\Delta C$ by the Erd\"os-R\'enyi value $C_{ER}=\langle k \rangle / N$. 
This $N$ dependence is not surprising, because the clustering coefficient itself vanishes at constant link density as $N\to \infty$. 
In summary, the maximum of the clustering coefficient is a possible robust warning sign for the upcoming persistence threshold.

%%%%%%%%%%%%%%%%%%%%%%%%%%%%%%%%%%%%%%%%%%%%%%%%%%%%%%%%%%%%%%%%%%%%%%%%%%%%%%%%%%%%%%%%%%%%
\subsection{Degree distribution}
\label{ssc:degr_distr}

\begin{figure}[t]
 \centering
% \hspace{-10ex}
\begin{tikzpicture}
% \node (empty) at (0,0) {};
\node[inner sep=0pt] (Fig1) at (0,0)
{\includegraphics[height=4.15cm]{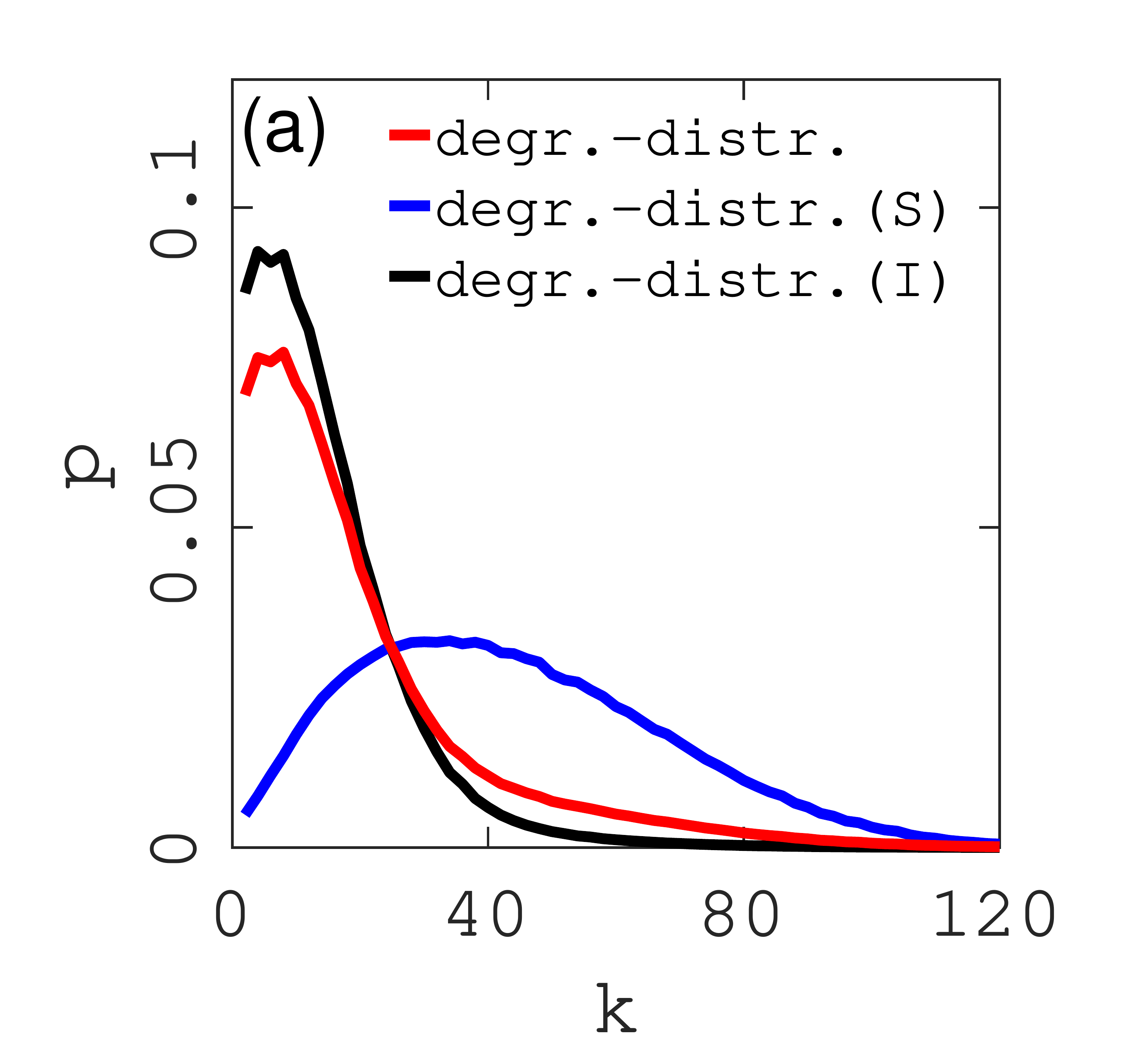}};
\node[inner sep=0pt] (Fig2) at (4.5,0)
{\includegraphics[height=4.15cm]{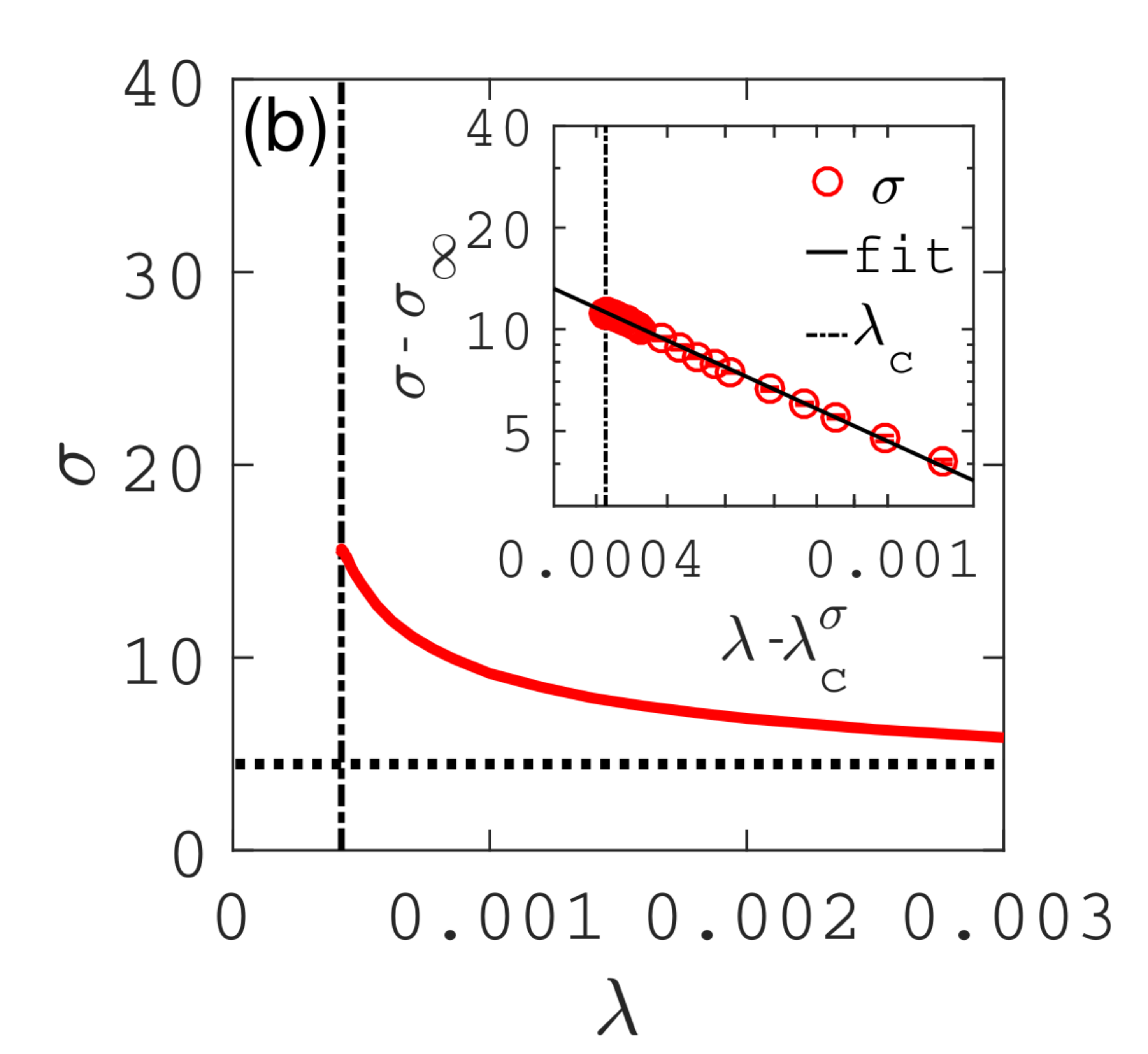}};
\end{tikzpicture}
  \caption{(a) Degree distribution of the entire graph (red), the susceptible nodes (blue), and the infected nodes (black) for $N=1000$, $\langle k\rangle=20$, $r=0.002$, $\lambda=0.001$, and $w=0.02$. 
 	The distributions are close to Poisson distributions as one would expect from ER networks. 
	(b) Standard deviation of the stationary degree distribution. The inset shows a power law fit. 
	The exponent is roughly $\beta_\sigma\approx -0.46$ and $\lambda_c^{\sigma}\approx 0.00022$ for $N=400$, 
	$\langle k \rangle=20$, $r=0.002$, $w=0.01$.
	}
  \label{fg:degree_distribution}
\end{figure}

The degree distribution $p_k$ is the fraction of nodes in the network with degree $k$. The average degree is 
$\langle k\rangle =1/N \sum_{i}k_i=2L / N$. Note that $\langle k \rangle$ is constant 
due to the conservation of links during rewiring. The $n$-th raw moment is given by 
\begin{equation}
 \langle k^n\rangle =\frac{1}{N}\sum_i k_i^n=\rho\langle k^n 
\rangle_I+\rho_S \langle k^n\rangle_S \quad,
\end{equation}
where $\langle k^n \rangle_{I(S)}$ are the raw moments 
of the degree distribution of infected (susceptible) nodes. 
The degree distribution in the endemic state has been studied for instance in~\cite{GrossDLimaBlasius,Marceauetal}. 
The degrees of the infected and the susceptible nodes both follow a Poisson distribution 
(an indication for ER random graphs), however with different mean values. 
The behavior in the vicinity of the phase transition at $\lambda_c$ has not been studied before. 

Figure~\ref{fg:degree_distribution}(a) shows the stationary degree distribution for a set of parameters close to the transition. 
The overall distribution is a superposition of the susceptible and the infected contribution. 
The respective Poisson distributions are seen. 
In Fig.~\ref{fg:degree_distribution}(b) we show the critical curve of the standard deviation $\sigma$ at equilibrium. 
We observe a rise of the standard deviation close to the critical point and the absence of a local maximum. 
For a Poisson distribution the mean $\langle k \rangle$ and variance $\sigma^2=\langle k^2\rangle-\langle k\rangle^2$ coincide, and 
under the assumption, that the infected and the susceptible nodes 
have a both a Poisson degree distribution, the overall variance is
\begin{equation}
{\rm Var} =\langle k \rangle + \left[  \rho \langle k \rangle_I^2+ (1-\rho)\langle k\rangle_S^2 
- \langle k\rangle^2 \right] \quad .
\end{equation}
Here we used $\langle k^2\rangle = \rho\langle k^2 \rangle_I+(1-\rho)\langle k^2 \rangle_S$
 and $\langle k^2\rangle_{I(S)}=\langle k\rangle^2_{I(S)}+\langle k\rangle_{I(S)}$.
By Jensen's inequality the term in square brackets is positive and we obtain the inequality, 
${\rm Var} \geq {\rm Var}_{\rm ER}= \langle k \rangle$. 
In Fig.~\ref{fg:degree_distribution}(b) the equilibrium standard deviation can be seen to be bounded from below by $\sigma_{\rm ER}=\sqrt{\rm Var_{\rm ER}}$, which is indicated by the dotted horizontal line. 

The inset in Fig.~\ref{fg:degree_distribution}(b) shows a power law fit. 
The curve is well described by a power law close to the transition, however, it's critical point is not {\em at} the transition. 
Like for the effective branching ratio (Fig.~\ref{fg:EBR}) the true critical point is not sensed by $\sigma$. 
% The values of the fitted parameters are very sensitive to the choice of the fit range. 
% This is likely due to the co-existence of multiple scaling laws. {\re I am not so sure about this???} 
The fitted critical points $\lambda_c^\sigma$ for various parameters and choices of intervals 
all share the feature that they are far away from $\lambda_c$ and close, or equal to zero. 
% {\re XXX}
% The fitted exponents vary between $-0.5$ and $-1.5$ depending on the choice of parameters and intervals. {\bl $\leftarrow$ WHAT SHOULD WE DO ABOUT THIS?}{\re I would delete the sentence from XXX on}

We conclude that the broadening of the degree distribution captures the approach towards the critical point, 
but the true location of the critical point $\lambda_c$ cannot be seen from the scaling behavior of $\sigma$.

%%%%%%%%%%%%%%%%%%%%%%%%%%%%%%%%%%%%%%%%%%%%%%%%%%%%%%%%%%%%%%%%%%%%%%%%%%%%%%%%%%%%%%%%%%%
\subsection{Degree assortativity}
\label{ssc:degass}

Assortativity measures the correlations between the degrees of adjacent nodes. 
In terms of the adjacency matrix $A$ and the degree vector $k_i=\sum_{j}A_{ij}$ it is 
\begin{equation}
 \mathcal A=\frac{\sum_{ij}\left(A_{ij}-\frac{k_ik_j}{N\langle k\rangle}\right)
k_ik_j}{\sum_{ij}\left(k_i\delta_{ij}-\frac{k_ik_j}{N\langle k\rangle}\right)k_ik_j} \quad .
\end{equation}
It takes values between $-1$ and $1$. For $\mathcal A=0$ the network has no 
degree correlations, for $\mathcal A=1$ it is maximally degree-correlated, and for 
$\mathcal A=-1$ it is maximally anti-correlated.

\begin{figure}[t]
 \centering
% \hspace{-10ex}
\begin{tikzpicture}
\node[inner sep=0pt] (Fig1) at (0,0)
{\includegraphics[height=4.4cm]{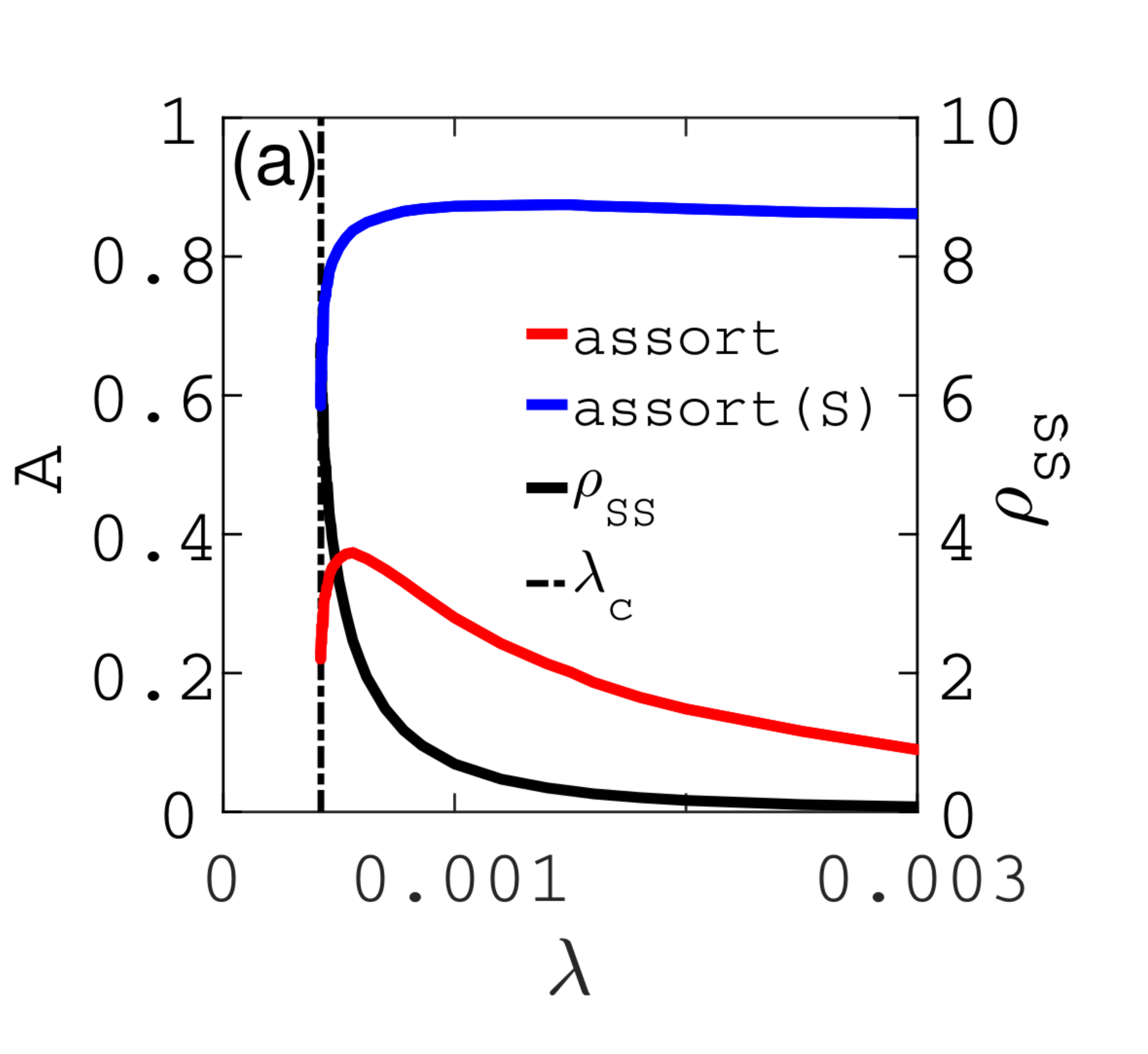}};
\node[inner sep=0pt] (Fig2) at (4.5,0.05)
{\includegraphics[height=4.0cm]{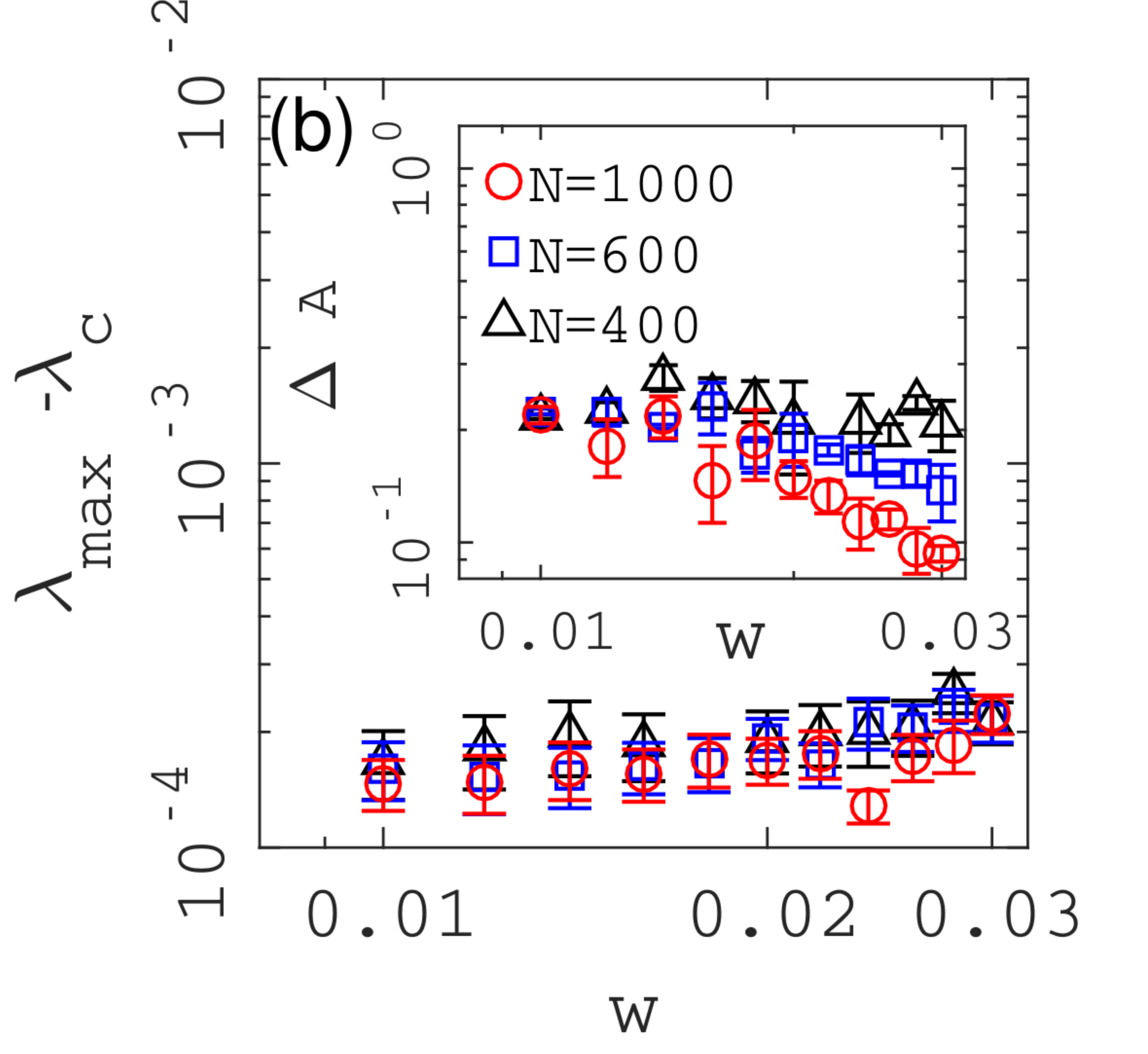}};
\end{tikzpicture}
  \caption{
	(a) Assortativity coefficient for the entire graph (red) and for the susceptible graph (blue). 
  	For reference we also show the density of $SS$ links (black), which approximates the size of the susceptible graph. 
	$N=400$, $\langle k \rangle = 20$, $r=0.002$, and $w=0.01$. 
	(b) The distance $\lambda_{\text{max}}$ and size $\Delta A$ (inset) of the local maxima shown for various $w$ and $N$. 
	%Parameters are the same.
	}
  \label{fg:assort_coeff}
\end{figure}

\begin{figure*}
\centering
% \hspace{-10ex}
\begin{tikzpicture}
\node[inner sep=0pt] (Fig1) at (0,0)
{\includegraphics[height=4.15cm]{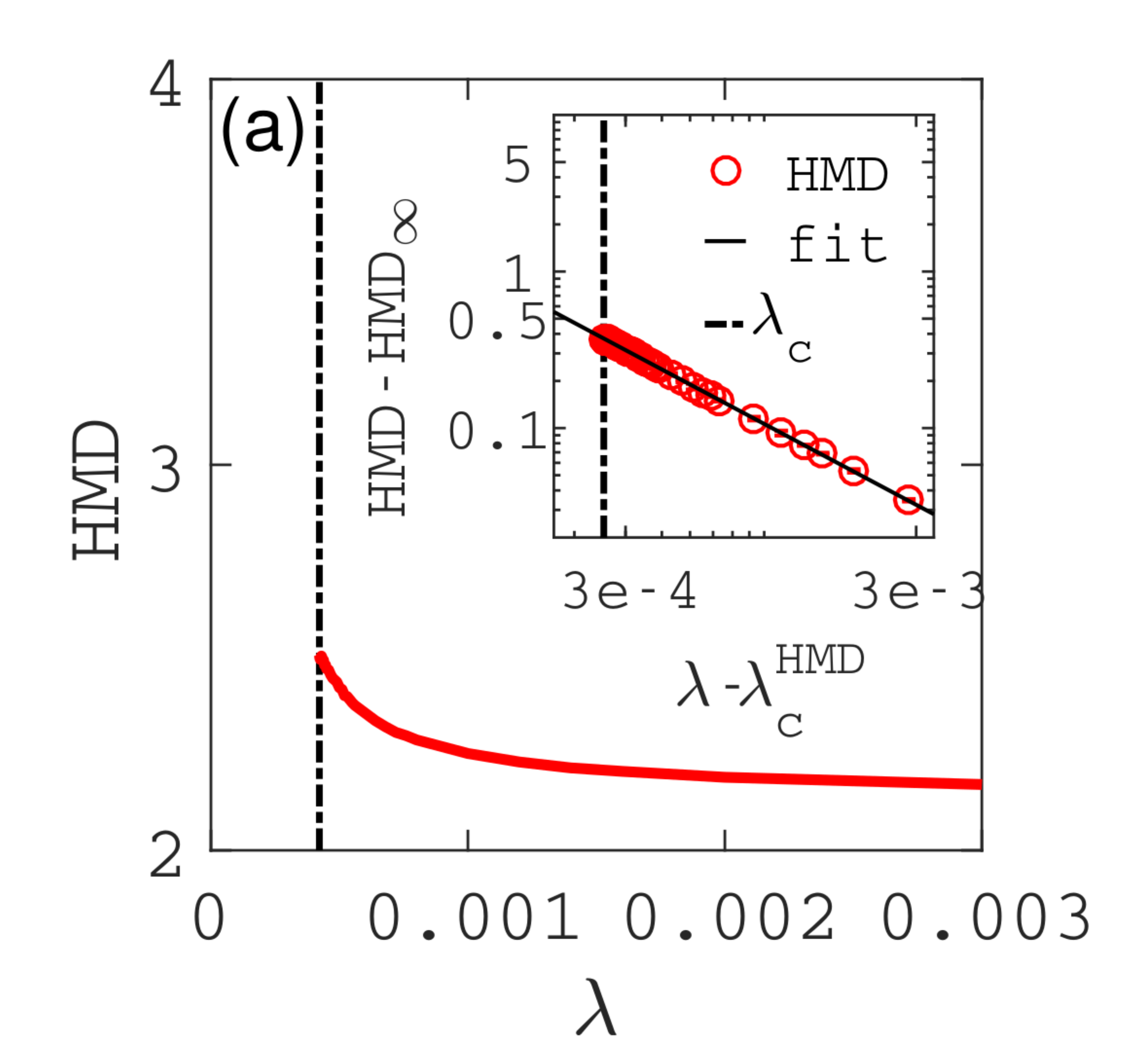}};
\node[inner sep=0pt] (Fig2) at (4.3,0)
{\includegraphics[height=4.15cm]{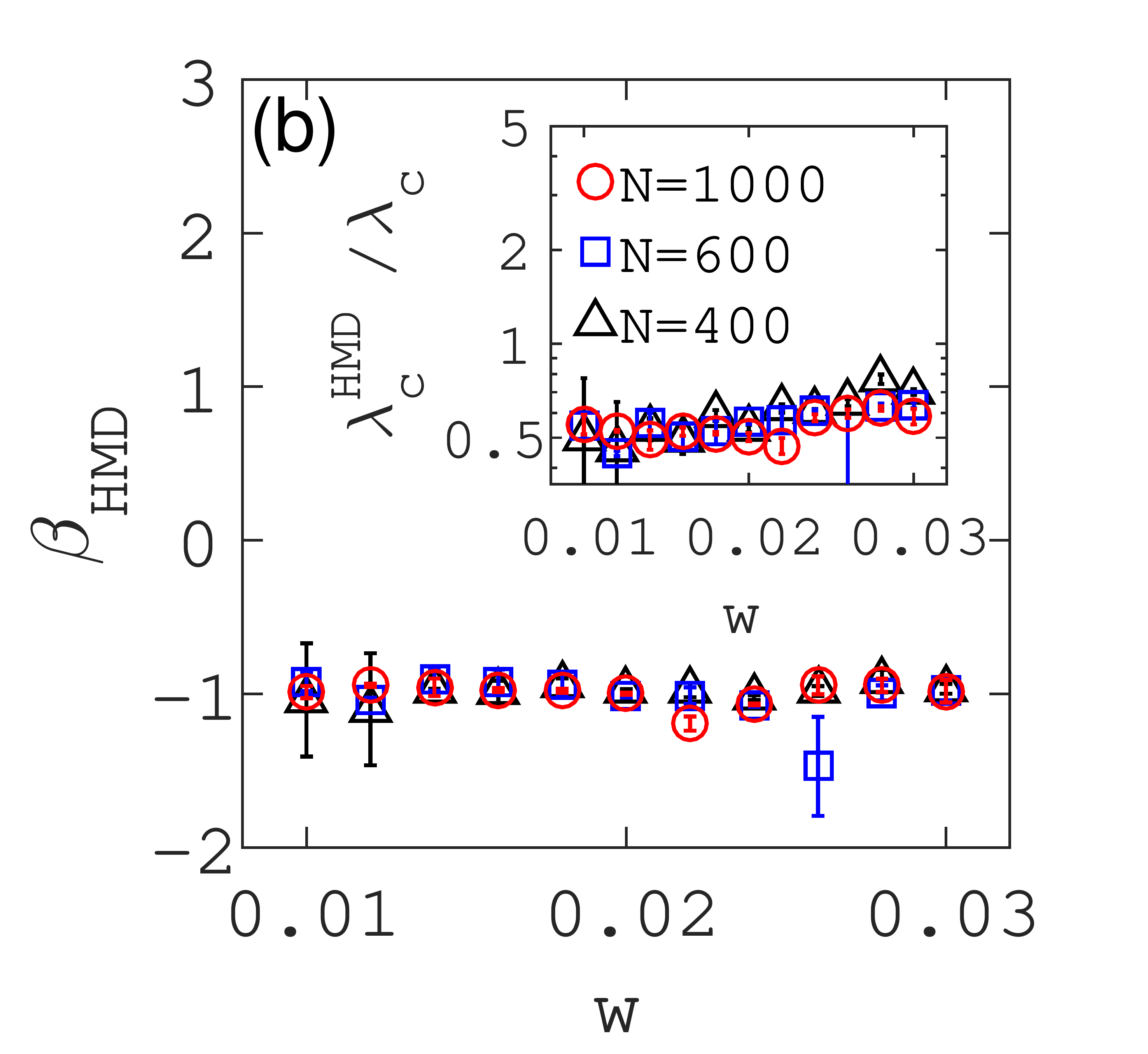}};
\node[inner sep=0pt] (Fig3) at (8.8,-0.15)
{\includegraphics[height=4.5cm]{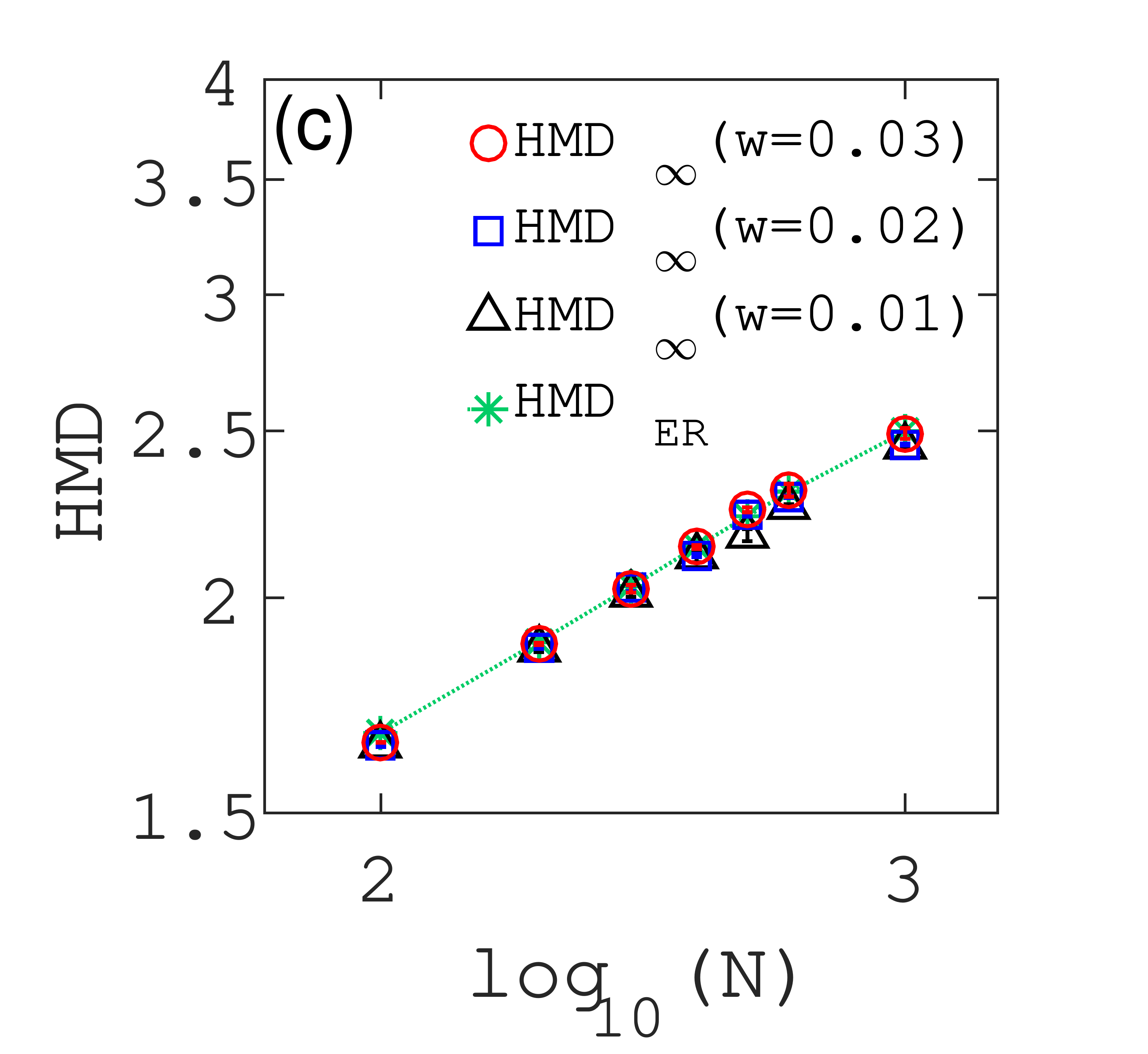}};
\node[inner sep=0pt] (Fig4) at (13.5,0)
{\includegraphics[height=4.15cm]{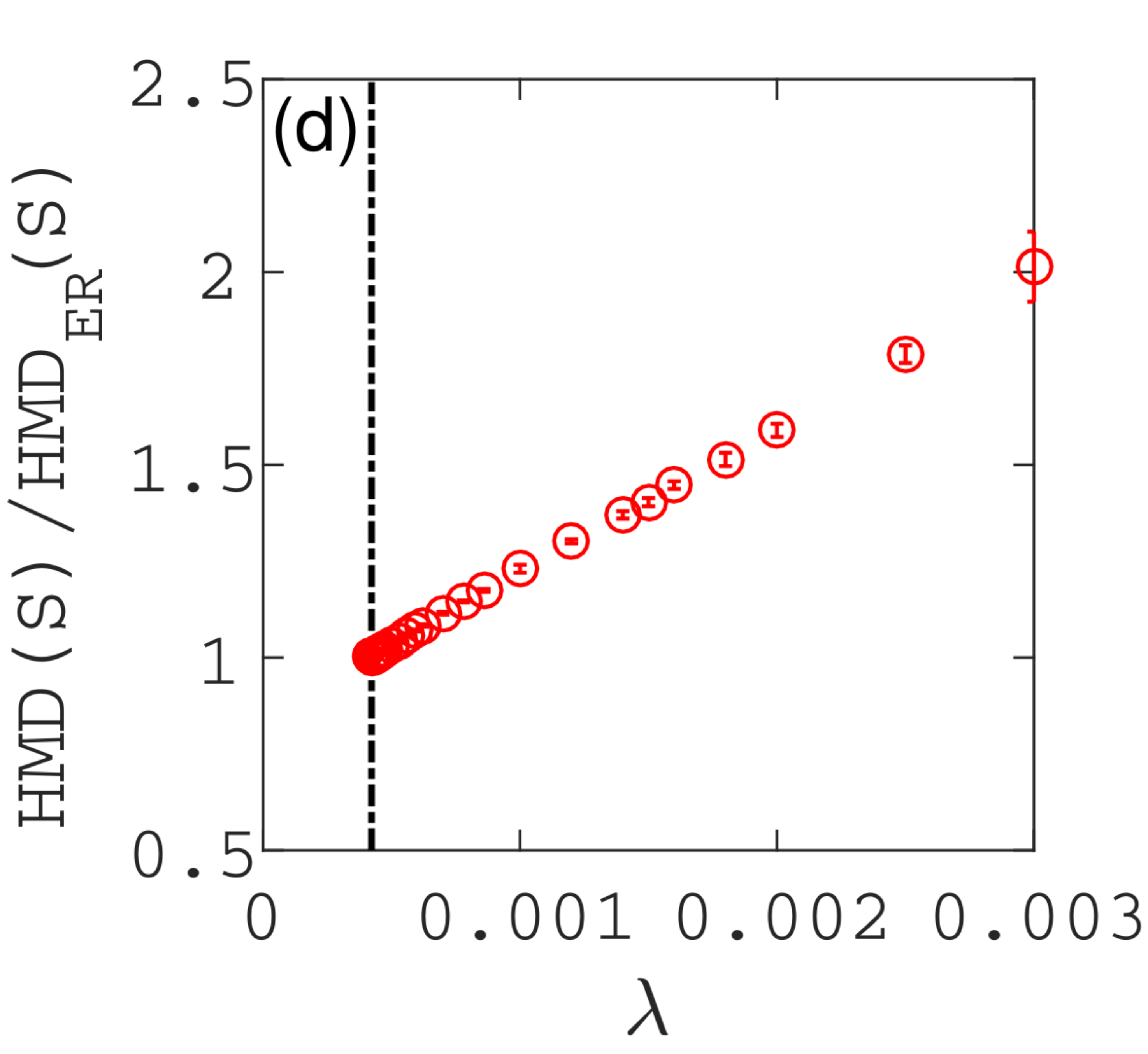}};
\end{tikzpicture}
  \caption{
  {(a) Harmonic mean distance $\rm{HMD}$ of the entire graph and a power law fit on a log-log plot (inset). 
  (b) Power law exponent $\beta^{\rm{HMD}}_c$ fitted to the tail and the critical point $\lambda_c^{\rm{HMD}}$ as a fraction of $\lambda_c$ (inset). 
  (c) Fitted asymptotic value ${\rm HMD}_\infty$ as a function of system size $N$, with the numerical values of the Erd\"os-R\'enyi ensemble average 
  ${\rm HMD}_{\rm ER}(N)$ for a reference. 
  (d) Harmonic mean distance of the susceptible subgraph $\rm{HMD}(S)$, measured with respect to the ER 
  value for a graph of that size $S=N (1-\rho)$. $N=400$, $\langle k\rangle=20$, $r=0.002$, and $w=0.01$}
	}
  \label{fg:harmonic_mean_distance}
\end{figure*}

Figure~\ref{fg:assort_coeff}(a) shows the critical curve for the assortativity. 
As for the $SI$ link density and the clustering coefficient, the degree assortativity exhibits a maximum. 
For large infection rates the network becomes non-assortative, which is the expected Erd\"os-R\'enyi limit. 
At medium-range infection rates the maximum occurs. 
Towards the threshold, the assortativity decreases again. 

It is instructive to decompose the assortativity coefficient into its constituent parts. 
We denote by $\langle k k' \rangle_{AB}$ the expected product of the degrees $k$ and $k'$ along links of type $AB$. 
In this notation the coefficient is
\begin{equation}
 \frac{2\rho_{SS}\langle kk'\rangle_{SS}+2\rho_{SI}
\langle kk'\rangle_{SI}+2\rho_{II}\langle kk'\rangle_{II}
-\frac{1}{\langle k \rangle}\langle k^2 \rangle^2}{ \langle
 k^3 \rangle-\frac{1}{\langle k \rangle}\langle k^2 \rangle^2} \quad .
\end{equation}
The important contribution to the overall assortativity results from the first three terms in the numerator. 
The assortativity of the susceptible subgraph rises to high values above $0.8$, as can be seen in Fig.~\ref{fg:assort_coeff} (a) 
This means that the most important contribution of the three terms comes from the susceptible subgraph. 
There is, however, a trade-off between the abundance of $SS$ links and the expected degree correlation 
$\langle kk' \rangle_{SS}$, while the latter is increasing, the former decreases (Fig.~\ref{fg:assort_coeff}(a)). 
The degree correlations, however, increase faster than the $SS$ link density decreases, thus giving rise to the maximum. 

The interference of at least two scaling laws brings about the maximum. 
Its location is studied in Fig.~\ref{fg:assort_coeff}(b). With respect to the distance of the maximum from the threshold $\lambda_{\text{max}}-\lambda_c$ one observes two small trends, namely a slight increase in distance towards higher rewiring rates, 
and a non-significant decrease with respect to system size. 
The height of the maximum $\Delta \mathcal A=\mathcal A(\lambda_{\text{max}})-\mathcal A(\lambda_c)$ (inset of Fig.~\ref{fg:assort_coeff}(b)) 
on the other hand has a strong size dependence. 
It decreases with increasing $w$, but does so at a higher rate as the system becomes larger. 
The decrease might follow a power law.

The degree assortativity is a purely global quantity that is not easy to measure locally. 
It does, however, bear potential as an early-warning sign because the distance to the threshold is 
sizable and not strongly dependent on the rewiring rate or system size.

%%%%%%%%%%%%%%%%%%%%%%%%%%%%%%%%%%%%%%%%%%%%%%%%%%%%%%%%%%%%%%%%%%%%%%%%%%%%%%%%%%%%%%%%%%%

\subsection{Harmonic mean distance}
\label{ssc:cmptness}

The most natural way of measuring distances on a graph is in terms of the geodesic distance. 
For two nodes $i$ and $j\neq i$ the geodesic distance $d_{ij}$ is the length of the shortest path between them, 
and is infinite if $i$ and $j$ are not connected by any path. 
The ``farness'' of a node $i$ is given by $1/(N-1)\sum_{j\neq i}d(i,j)$, the ``closeness'' by its reciprocal. 
For graphs with multiple connected components the farness is infinite and the closeness vanishes. 
To remedy this deficiency one may look at the harmonic geodesic distance $1/d_{ij}$. The harmonic mean geodesic 
distance is then given by
\begin{equation}
{\rm HMD} =\frac{1}{\langle \frac{1}{d}\rangle}=\frac{N(N-1)}{\sum_{i,j\neq i}\frac{1}{d_{ij}}} \quad .
\end{equation}
It is $1$ for a complete graph, infinite for a set of points without links, and finite otherwise.

\begin{figure*}
\centering
% \hspace{-10ex}
\begin{tikzpicture}
% \node (empty) at (0,0) {};
\node[inner sep=0pt] (Fig1) at (0,0)
{\includegraphics[height=4.15cm]{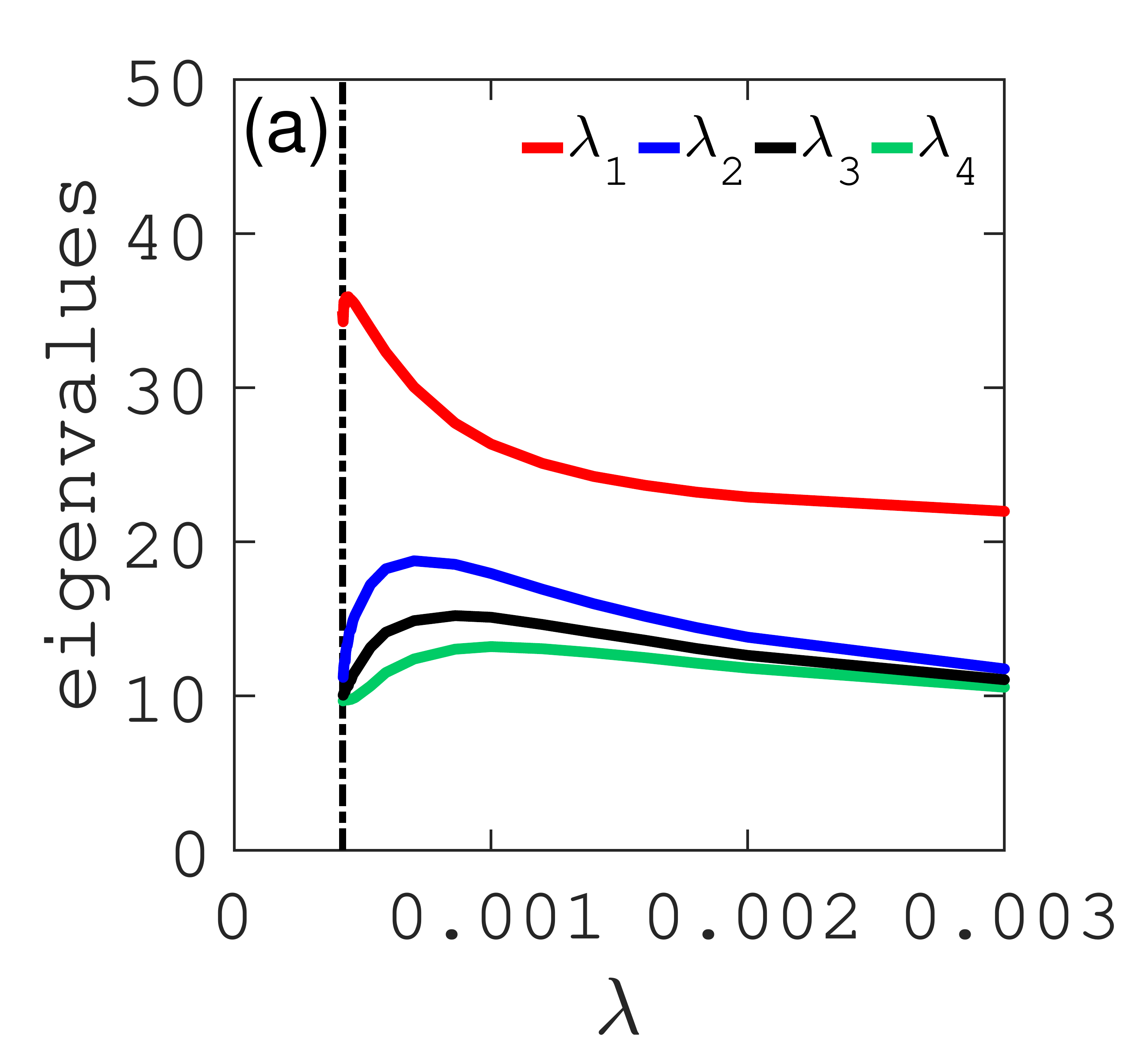}};
\node[inner sep=0pt] (Fig2) at (4.5,0)
% {\includegraphics[width=0.251\linewidth]{finalpics/average_local_clustering_gimped.pdf}};
{\includegraphics[height=4.15cm]{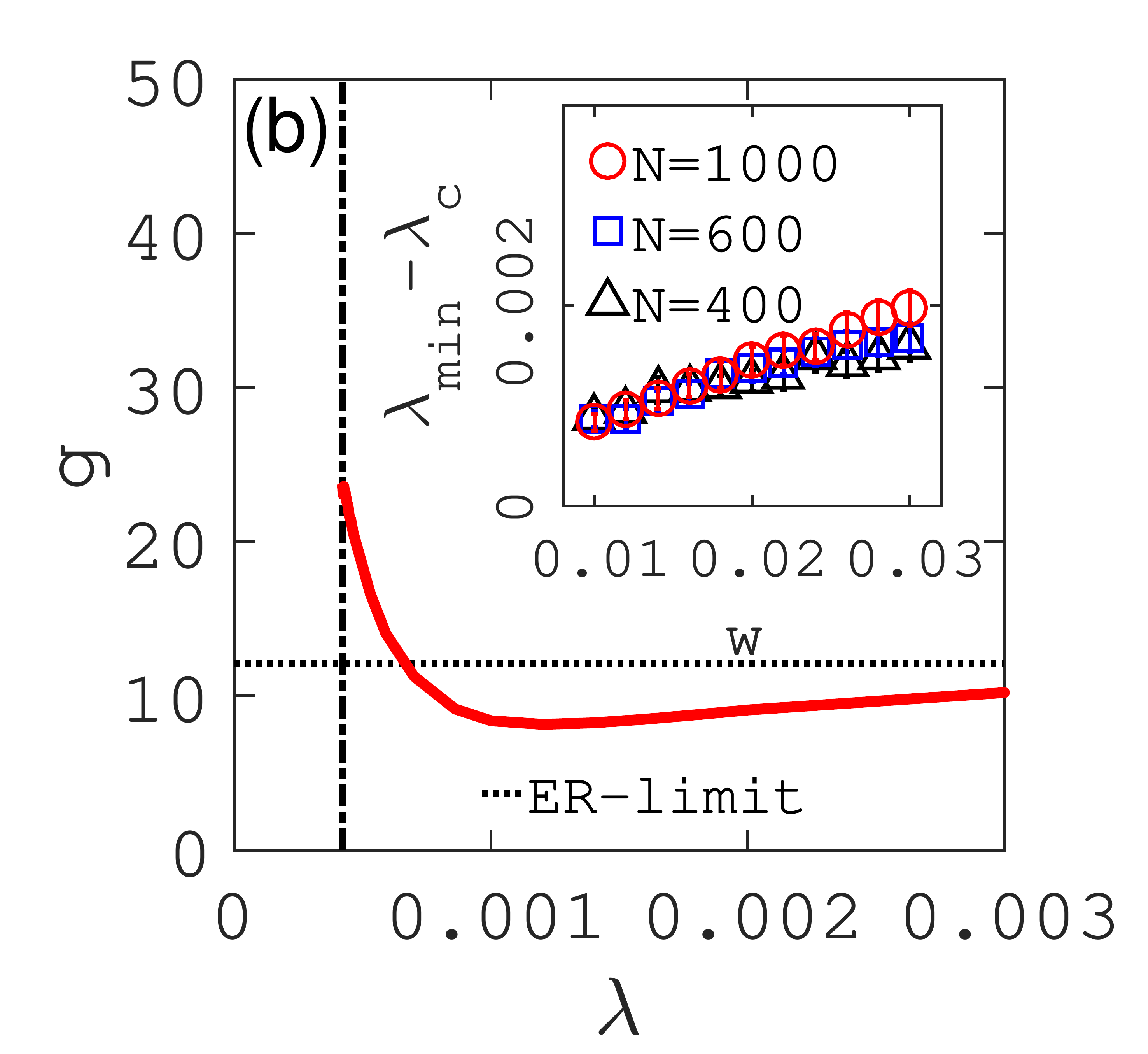}};
\node[inner sep=0pt] (Fig3) at (9,0)
{\includegraphics[height=4.15cm]{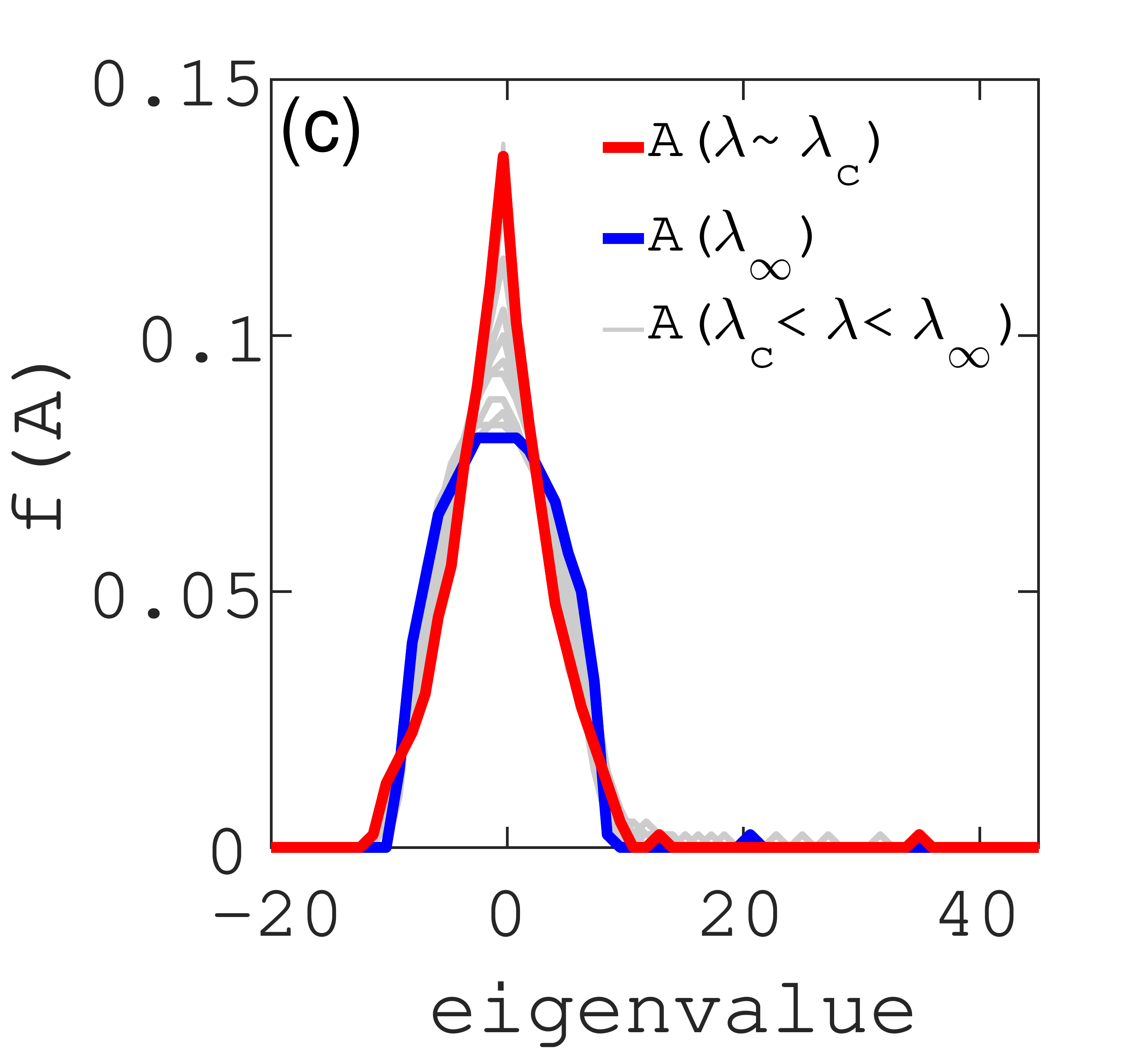}};
\node[inner sep=0pt] (Fig4) at (13.5,0)
% {\includegraphics[width=0.251\linewidth]{finalpics/cl_distances_inset_gimped.pdf}};
{\includegraphics[height=4.15cm]{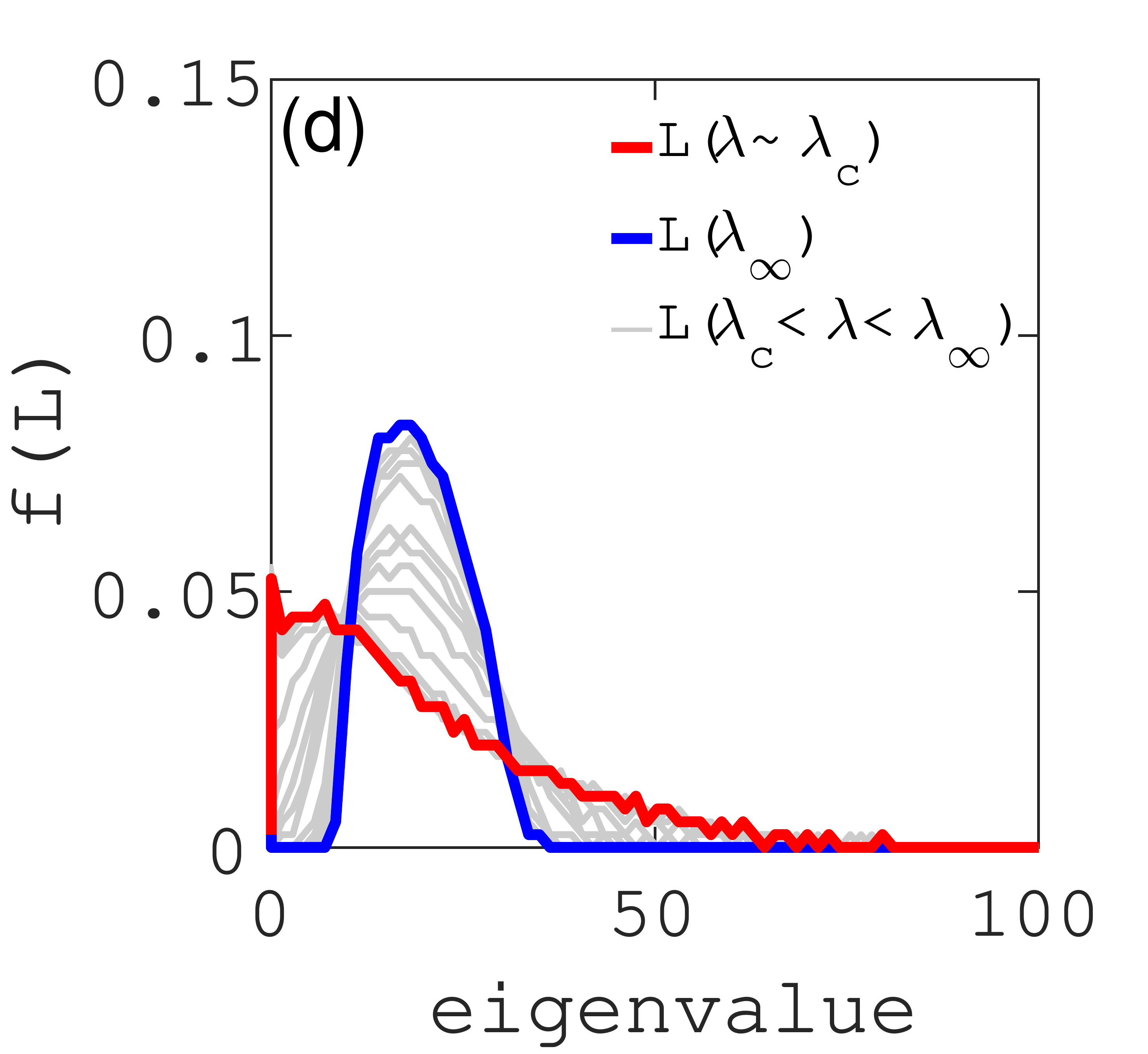}};
% \node[inner sep=0pt] (Fig3) at (0,0)
% {\includegraphics[width=.55\linewidth]{finalpics/TimeSeries_wlegend.pdf}};
% \node[inner sep=0pt](a) at (0.2,-2.3){(a)};
% \node[inner sep=0pt](b) at (4.72,-2.3){(b)};
% \node[inner sep=0pt](c) at (9.36,-2.3){(c)};
% \node[inner sep=0pt](d) at (13.86,-2.3){(d)};
% \node[draw,dotted,inner sep=2mm,label=below:\textb{fSome comment below picture} (Section ),fit=(Fig1) (Fig2) (Fig3)] {};
\end{tikzpicture}
  \caption{
  (a) The first 4 eigenvalues. 
  (b) The eigenvalue gap and the distances from the threshold to the minimum for a range of rewiring rates and system sizes (inset). 
  (c) Distribution of eigenvalues of the adjacency matrix, and 
  (d) the network Laplacian.
	Blue curves correspond to high values of the infection rate, $\lambda=0.03$, and show Wigner's semi circle.
	Red lines correspond to the distributions for infection rates close to the critical threshold, $\lambda_c=0.00043$. 
	Grey lines show distributions for  intermediate infection rates. 
	$N=400$, $\langle k\rangle=20$, $r=0.002$, and $w=0.01$. 
	}
  \label{fg:eig_adj}
\end{figure*}

Figure~\ref{fg:harmonic_mean_distance}(a) shows the ${\rm HMD}$.
For infection rates close to the critical point, the ${\rm HMD}$ rises--the network becomes less compact. 
A possible explanation is that a large number of paths in the network lead through the susceptible subgraph, 
especially for infected nodes. 
This is supported by the high branching ratio $\kappa=[SSI]/[SI]$ close to the critical point. 
Therefore, the overall distances become larger in the vicinity to the critical threshold. 
For large infection rates the equilibrium distances approach the ensemble average of the Erd\"os-R\'enyi 
harmonic mean distance $\rm{HMD}_{ER}\sim \log(N)$, as seen in Fig.~\ref{fg:harmonic_mean_distance}(c). 
The susceptible subgraph itself is at its densest near the threshold and its ${\rm HMD}$ in comparison to that of an 
ER graph of the same size grows linearly (Fig.~\ref{fg:harmonic_mean_distance}(d)). 
An explanation is that the number of $SS$ links falls by a factor of $\lambda^{-1}$ faster than the number of susceptible nodes $S=N(1-\rho)$. 
So the average distances of the susceptible subgraph scale by the reciprocal factor with respect to a baseline ER graph of size $S$. 

Like for the effective branching ratio $\kappa$ and for the standard deviation $\sigma$  (Subsections \ref{ssc:ebr} and
\ref{ssc:degr_distr}), the ${\rm HMD}$ does not sense the actual critical point $\lambda_c$. 
A fit to a power law reveals that the exponent is close to $-1$ for a range of rewiring rates and system sizes (Fig.~\ref{fg:harmonic_mean_distance}(b)). The fitted critical point $\lambda_c^{\rm HMD}$ is both bigger than $0$ and strictly smaller than $\lambda_c$.

%%%%%%%%%%%%%%%%%%%%%%%%%%%%%%%%%%%%%%%%%%%%%%%%%%%%%%%%%%%%%%%%%%%%%%%%%%%%%%%%%%%%%%%%%%%%%%%%%%
\subsection{Spectral properties}

The network Laplacian is defined in terms of the adjacency matrix $A$ by $L=D-A$, with $D_{ij}=\delta_{ij} \sum_k A_{jk}$
The largest eigenvalue of the adjacency matrix, $\lambda_{1}$, is sometimes referred to as the ``capacity'' of the graph. 
It carries information about the connectivity and the number of paths in the network. 
It is bounded from below by the average degree $\langle k\rangle$, and from above by the maximal degree. We denote the gap between the first two eigenvalues by $g=\lambda_1-\lambda_2$. In the context of Markov chains it measures the speed of convergence in $\ell^p$ to the stationary distribution under the condition of irreducibility and aperiodicity, 
see \cite{LevinPeres,Chung,ChungLu} for more details. 

Figure~\ref{fg:eig_adj}(a) shows the five largest eigenvalues of the adjacency matrix. 
The largest eigenvalue, unlike the remaining ones, attains a maximum shortly after the threshold and then decreases towards an asymptotic value. 
In Fig.~\ref{fg:eig_adj}(b) we see that the eigenvalue gap is large when close to the threshold, drops steeply to a local minimum, 
and then relaxes back to its asymptotic value. 
The larger the gap the more difficult it is to dissect the graph \cite{ChungLu} and the faster infections would spread. 
The distance of the local minimum scales linear with the rewiring rate, as can be seen from the inset of Fig.~\ref{fg:eig_adj}(b).
Therefore it becomes a very reliable indicator of the transition and even more so, for higher rewiring rates.

Another important spectral characteristic of the network is the distribution of eigenvalues. 
It is known~\cite{FurediKomlos} that the empirical eigenvalue distribution 
of the adjacency matrix converges to the Wigner semi-circle law for ER graphs. 
The Laplacian, however, converges to the convolution of a Gaussian with the semi-circle distribution~\cite{DingJiang}, 
after appropriate normalisation. 
Figure~\ref{fg:eig_adj} shows the empirical distribution of both the adjacency matrix (c) and the graph Laplacian (d). 
Far from the threshold, the eigenvalue distribution of the adjacency matrix approaches the semi-circle around the origin, as expected. 
Close to the critical threshold the distribution changes drastically. 
It remains symmetric around the origin, but develops a narrow peak producing a cusp at the center. 
This behavior is known from eigenvalue distributions of several scale-free networks~\cite{ChungLu}. 
For the empirical eigenvalue distribution of the Laplacian the situation is similar: an
ER limit exists for high infection. Drastic changes occur near the critical threshold.

%%%%%%%%%%%%%%%%%%%%%%%%%%%%%%%%%%%%%%%%%%%%%%%%%%%%%%%%%%%%%%%%%%%%%%%%%%%%%%%%%%%%%%%%%%%
\section{Discussion -- usability of network measures as early-warning signs}
\label{sc:powerlove}

In summary, we formulated the general question of the feasibility of finding network-based precursor signals in adaptive network dynamics 
in the context of a specific epidemic model, the co-evolving SIS model. 
We find that several network measures indicate no sensitivity whatsoever, for the critical transition.
These are the effective branching ratio, the degree distribution and the harmonic mean distance. 
As a function of the infection rate, these measures show scaling laws, often characterized by an exponent of $-1$, 
with the singularity located at zero or close to it, but {\em not} at $\lambda_c$. 
It means that these measures behave as if the transition was at $\lambda^{\rm ebr / degree / HMD}_c  \approx 0$, 
$\lambda^{\rm ebr / degree / HMD}_c  \ll \lambda_c$,
rather than at $\lambda_c$. This has severe consequences for their use as an early warning sign, 
because the fold bifurcation point is suddenly reached without any warning. 
These measures can in no way anticipate the true position of the critical point $\lambda_c$.

We have shown, however, that a number of other network measures do carry potential for being used as early-warning signs. 
They are able to detect the critical transition at $\lambda_c$, when approached from above ($\lambda > \lambda_c$).
In particular, these measures, which include the
$SI$ link densities, 
triplet densities, 
clustering, and 
assortativity, 
show a crossover of two scaling laws, that are of a functional form as in Eq.~\eqref{eq:model}. 
The first scaling law shows an increase of the respective measure as $(\lambda-\lambda_c)^{1/2}$, 
slightly above the transition ($\lambda > \lambda_c$). 
The other is an asymptotic scaling law ($\lambda \gg \lambda_c$), which is characterized by negative integer exponents. 
Between these two scaling regimes a local maximum exists, which is indeed visible in the corresponding network measures. 
The location of the maxima occur slightly above the critical point, $\lambda^{\rm max} > \lambda_c$.

Both, the double scaling and the maximum is also seen in the maximum eigenvalue of the adjacency matrix, 
when plotted against $\lambda$. 
The eigenvalue gap shows a very clear minimum, well before the critical transition. 
In practical terms this means that, when approaching the critical transition point, an increase of the 
eigenvalue gap signals the immediate vicinity of the transition. 
Given a sufficiently robust eigenvalue estimate from data, the eigenvalue gap is a very clear and practical early-warning sign. 

We tested the effects of all parameters in the co-evolving SIS model and found that our results are relatively robust. 
The dependence on rewiring rate and system size has been has been investigated especially carefully. 
The recovery rate sets the time scale and can therefore can be fixed arbitrarily; we took the choice used in \cite{GrossDLimaBlasius}. 
The connectivity determines the location of the threshold. 
The homogeneous pair approximation  becomes unreliable for very low values of $\langle k \rangle$ \cite{Marceauetal}.

We conclude by mentioning that the network information that is neglected in the classical 
coarse-graining approach does contain a layer of structural information, that can indeed detect the critical transition point. 
The next steps to take would be to actually test the performance of the different network measures as precursor signals 
in agent based simulations, where infection rates are exogenously varied slowly. 
An observer can monitor the networks, the infection-, and rewiring rates, but would not know anything about the location of the critical point. 
It would be interesting to see to what extent such an observer could predict the collapse of the 
system several timesteps in advance. 

Supported by the Austrian Science Foundation FWF under the project P29252. 

%%%%%%%%%%%

\bibliographystyle{plain}
% \bibliography{AdaptiveSISNearCriticality_article_27Jan}

% \input{AdaptiveSISNearCriticality.bbl}

\begin{appendix}

%%%%%%%%%%%%%%%%%%%%%%%%%%%%%%%%%%%%%%%%%%%%%%%%%%%%%%%%%%%%%%%%%%%%%%%%%%%%%%%%%%%%%%%%%%%
\vspace{3ex}
\section{Quality of the Pair Approximation}
\label{ssc:PA}

In simulations the quality of the homogeneous pair approximation can be measured directly 
through the ratio of the exact triplet density over the approximate triplet densities, 
\be
\label{eq:tripletapprox1}
\varepsilon_{SSI}=\frac{\rho_{SSI}}{2\rho_{SI}\rho_{SS}/\rho_{S}}, \quad 
\varepsilon_{ISI}=\frac{\rho_{ISI}}{\rho_{SI}\rho_{SI}/\rho_{S}} \quad .
\ee

\begin{figure}[h!]
 \centering
% \hspace{-10ex}
\begin{tikzpicture}
% \node (empty) at (0,0) {};
\node[inner sep=0pt] (Fig1) at (0,0)
{\includegraphics[height=4.15cm]{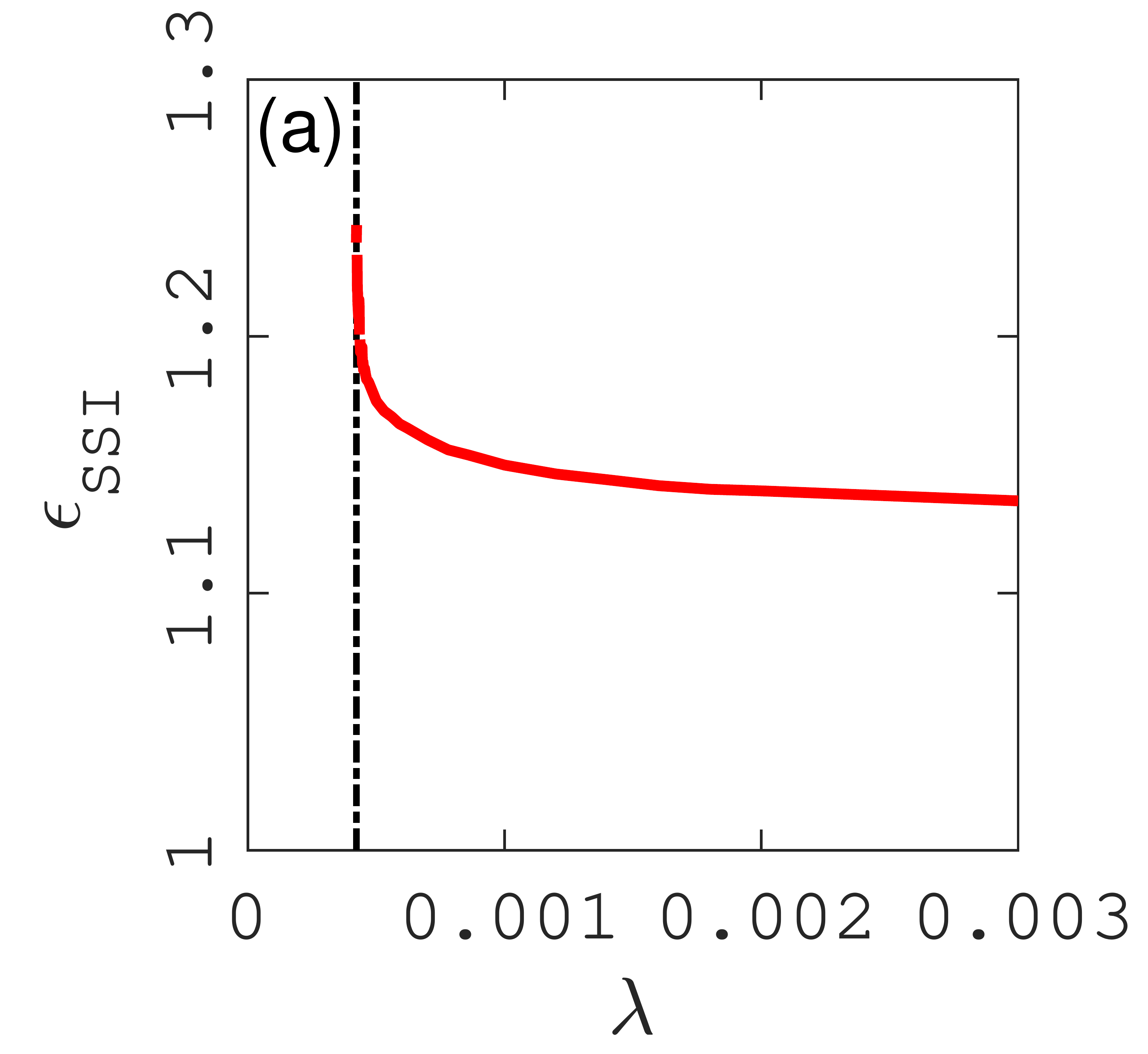}};
\node[inner sep=0pt] (Fig2) at (4.5,0)
{\includegraphics[height=4.15cm]{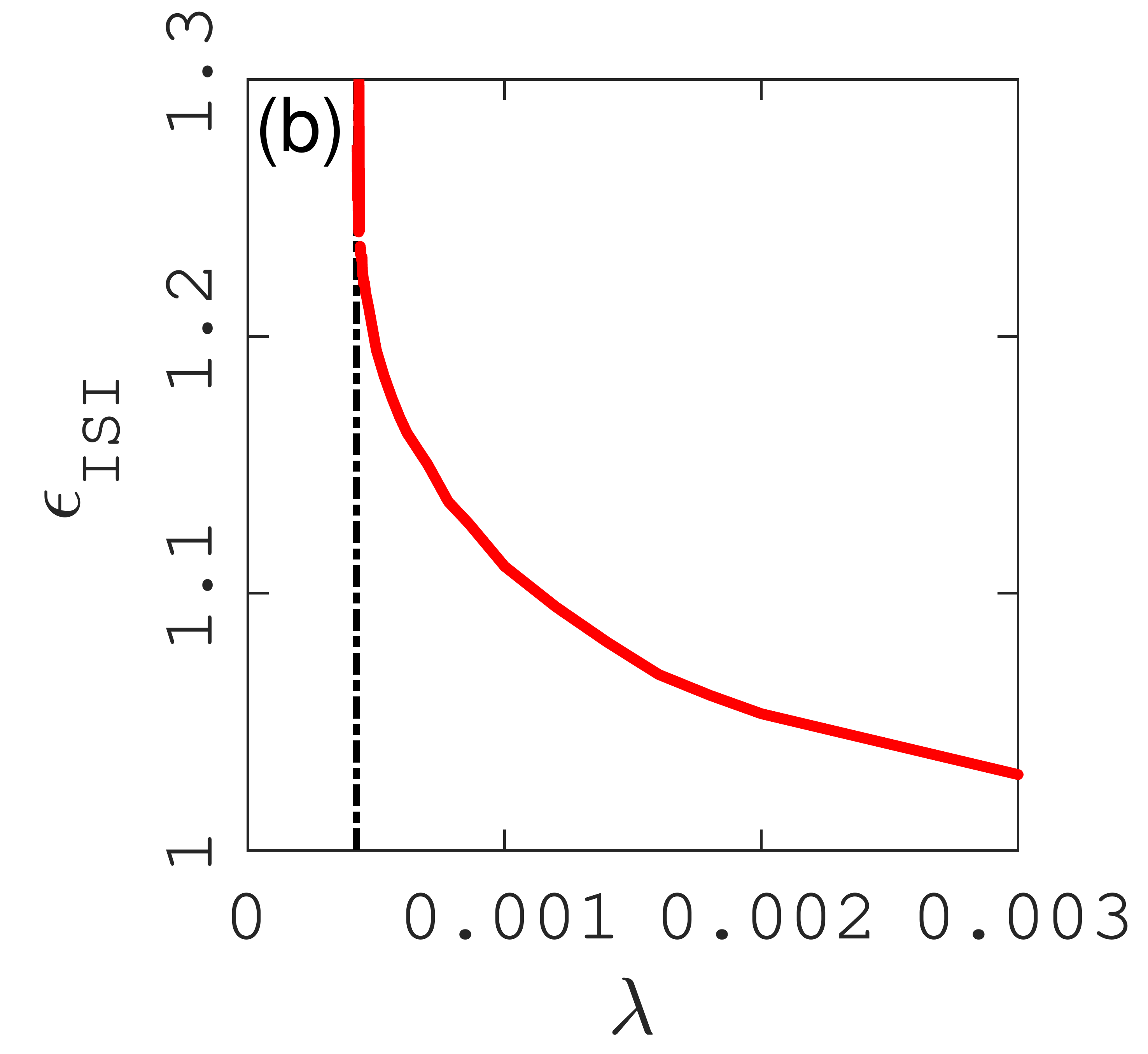}};
\end{tikzpicture}
  \caption{
	Ratio of the exact triplet densities and their pair approximation is 
	plotted at stationarity (a) for $\varepsilon_{SSI}$ and (b) $\varepsilon_{ISI}$.
	$N=400$, $\langle k \rangle=20$, $r=0.002$, and $w=0.01$.
	}
  \label{fg:pairapprox}
\end{figure}

For ratios above one the approximation underestimates the true triplet density.
Figure~\ref{fg:pairapprox} indicates that the quasi-stationary density of $SSI$ and $ISI$ 
triplets is systematically underestimated, which confirms the conjecture that moment closure 
becomes problematic near the instability~\cite{KuehnMC}. 
Further, it shows that correlations between different moment orders indicate proximity to the persistence threshold. 

\end{appendix}

\end{document}